# Flux effects in precipitation under irradiation – simulation of Fe-Cr alloys

Jia-Hong Ke [a], Elaina R. Reese [b], Emmanuelle A. Marquis [b], G. Robert Odette [c], and Dane Morgan [a*]

[a] Department of Materials Science and Engineering, University of Wisconsin-Madison, Madison, WI 53706, USA
[b] Department of Materials Science and Engineering, University of Michigan, Ann Arbor, MI 48109, USA
[c] Department of Materials Science and Engineering, University of California, Santa Barbara, CA 93106, USA
* Corresponding author. E-mail address: ddmorgan@wisc.edu

## Abstract

Radiation-enhanced precipitation of Cr-rich α' in irradiated Fe-Cr alloys, which results in hardening and embrittlement, depends on the irradiating particle and the displacement per atom (dpa) rate. Here, we utilize a Cahn-Hilliard phase-field based approach, that includes simple models for nucleation, irradiating particle and rate dependent radiation-enhanced diffusion and cascade mixing to simulate α' evolution under neutrons, heavy ions, and electron irradiations. Different irradiating particles manifest very different cascade mixing efficiencies. The model was calibrated using neutron data. For cascade inducing neutron/heavy-ion dpa rates at 300 °C between $10^{-8}$ and $10^{-6}$ dpa/s the model predicts approximately constant number density, decreasing radius, decreasing α' Cr composition, and lower α' volume fraction. The model then predicts a dramatic transition to no α' formation above approximately $10^{-5}$ dpa/s, while electron irradiation, with weak mixing, had little effect at dpa rates up to $10^{-3}$ dpa/s. These model predictions are consistent with experiments. We explain the results in terms of the flux dependence of the radiation enhanced diffusion, cascade mixing, and their ratio, which all vary significantly in relevant flux ranges for neutron and cascade inducing ion irradiations. These results show that both cascade mixing and radiation enhanced diffusion must be accounted for when attempting to emulate neutron-irradiation effects using accelerated ion irradiations.

# 1 INTRODUCTION

The stability and irradiation tolerance of structural materials is a critical factor in determining the lifespan and safety of nuclear power plants. For example, precipitation hardening and embrittlement, enhanced or induced by in-reactor neutron irradiation, is a major concern. There is a significant body of research on irradiation-enhanced precipitation that includes highly accelerated heavy ion and energetic electron studies. However, the use of ions and electron data to infer and/or extrapolate in service neutron irradiation behavior requires a quantitative understanding of both dose rate and irradiating particle effects. Note that in this paper we will refer to any ion that induces cascades as a heavy ion (including protons), to distinguish these cascade-inducing ions from non-cascade-inducing electrons.

One example where both dose rate and irradiating particle effects are observed is α' (Cr-rich bcc ferrite) precipitation in high-chromium steels. These alloys are of particular interest for reduced-activation ferritic/martensitic (F/M) steels and nanostructured ferritic alloys containing 8-14 wt.% Cr that are candidates for advanced fission (Generation IV) and fusion reactor applications. Early studies showed that α' formed under thermal annealing at temperatures between 350 and 550 °C and resulted in the so-called "475 °C embrittlement" in steels containing >14% Cr [1–3]. However, at lower temperatures (~300 °C), even though the driving force of phase separation is larger, α' precipitates were not observed under thermal aging because of sluggish diffusion [4]. In contrast, accelerated precipitation of α', which is responsible for irradiation embrittlement, was reported in neutron-irradiated F/M steels and Fe-Cr model alloys containing 10% Cr at 420 °C [5] and 13% Cr at 380 °C [6]. Small-angle neutron scattering (SANS) measurements by Mathon *et al.* showed the formation of α' in neutron-irradiated F/M commercial steels containing ≥ ~9% Cr at even lower temperatures of 250 and 325 °C [7]. Precipitation of α' at 300 °C has also been observed in neutron irradiations of Fe-Cr model alloys containing ≥ 9% Cr, based on atom probe tomography (APT) and SANS characterization techniques [8–16]. Alternatively, Horton *et al.* did not observe α' precipitation in a Fe-10Cr irradiated by triple-beam ions at 450 °C to 10 dpa [17]. Pareige *et al.* recently reported an APT study of Fe-12Cr under heavy-ion irradiation at $1\times10^{-4}$ dpa/s at 300 °C that observed a low density of dilute Cr clusters after irradiation to 0.5 dpa [18]. Likewise, Marquis *et al.* [19] found no α' in Fe-15Cr irradiated by $Fe^{++}$ ions at a dose rate of $1\times10^{-4}$ dpa/s up to 60 dpa at 300 °C, but observed a small density clusters with only 35-50 at.% Cr in in a Fe-15Cr at a lower dose rate of $\approx 1\times10^{-5}$ dpa/s, suggesting a pronounced dose rate effect [19]. Korchuganova *et al.* carried out a heavy-ion irradiation at room temperature on a Fe-22Cr alloy that was pre-aged at 500 °C to form α' precipitates with a distribution of sizes [20]. APT showed that the heavy-ion irradiation dissolved the small preexisting α' precipitates and made the larger ones more diffuse. A similar effect of heavy-ion irradiation was also shown to retard spinodal decomposition of Cr-rich α' in a duplex stainless steel with 20 wt.% Cr at 300 °C [21]. In contrast to the heavy-ion irradiation microstructures, Tissot *et al.* reported that a $3.9\times10^{-5}$ dpa/s electron irradiation at 300 °C accelerated precipitation of near equilibrium α' [22].

Accelerated α' precipitation under neutron irradiation has been attributed to radiation-enhanced diffusion (RED). RED depends on both temperature and dose rate due to the corresponding effect on vacancy and self-interstitial atom recombination mediated concentrations through a variety of mechanisms. However, a dose rate effect on RED does not fully explain the observed differences in the numbers, sizes and compositions on Cr-enriched clusters in ion irradiations. We hypothesize that the one mechanism responsible for such dose rate effects on α' formation is cascade mixing, due to atomic replacements/relocations in displacement cascades. This so-called forced cascade mixing destabilizes α' precipitates by driving Cr atoms back into solution; hence, precipitation behavior is determined by the competition between cascade mixing and kinetics of precipitation. Indeed, a large number of studies have shown that cascade mixing can cause precipitate phase instabilities of various types, including radiation-induced re-resolution and disordering, inverse coarsening and re-precipitation [23–28].

Note that cascade mixing scales linearly with the cascade generation rate, while the thermodynamically driven Cr clustering occurs at rates controlled by RED, which has less than linear scaling with the flux when recombination is important. Therefore, we expect that for a high enough dose rate, cascade mixing will dominate and destabilize the precipitates. Furthermore, cascade mixing becomes more dominant at low temperature relative to the thermal diffusion and vacancy mobility, since temperature is expected to have almost no effect on cascade mixing. Note that although low temperature can also increase the defect concentration which increases the radiation enhancement to the kinetics, this effect is relatively small compared to the reduced thermal diffusion at lower temperature, so the overall RED is reduced. Irradiation with significant cascade mixing can ultimately result in a steady precipitate composition state that depends on the flux, temperature and damaging particle, and the type of displacement events it produces. If cascade mixing is dominant, the system never reaches thermodynamic equilibrium and the phase boundaries are modified by irradiation [25,26,29]

There is no consistent understanding in the literature regarding the cascade mixing effect on α'. Vörtler *et al*. carried out molecular dynamics (MD) simulations showing that pre-existing 5 nm diameter Cr-rich clusters are not dissolved or modified by 20 keV cascades in Fe-10Cr at 300 K, partly due to cascade branching [30]. MD studies by Tikhonchev *et al*. employed an improved N-body interatomic potential and modeled Cr-rich clusters with various diameters, showing that displacement cascades also produced only minor effects on Cr-rich clusters with diameters ≥ 3 nm, while small 1-2 nm clusters dissolved at 300 K [31]. Soisson and Jourdan [32] reported Atomistic Kinetic Monte Carlo (AKMC) simulations that showed no significant influence of cascade mixing on radiation-enhanced α' precipitation at 290 °C, even at high heavy-ion irradiation dose rates. Note, the effects of cascade mixing are much stronger for high-energy cascades produced by neutrons and heavy ions, versus the vacancy self-interstitial Frenkel pair defects produced by high-energy electrons.

In the present study, we employ Cahn-Hilliard-type phase-field simulations combined with nucleation, RED, and cascade mixing models to investigate α' precipitation under heavy ion, neutron, and high energy electron irradiation over a wide range of dose rates at two temperatures. Such a hybrid phase-field approach has been successfully applied to study nucleation and growth for various solid-state phase transformations [33–38], as well as simulating α' precipitation in Fe-Cr alloys under thermal annealing [39,40]. However, no previous phase-field study treated the additional coupled effects of RED and cascade mixing. We incorporated the continuum treatment of Enrique and Bellon [41], which was previously used to study compositional patterning and phase stability under irradiation [41–46] into our hybrid α' precipitation model. The hybrid phase-field formulation takes full advantage of treating concurrent diffusional and ballistic events from a continuum perspective. The model of concurrent diffusional clustering and cascade mixing is used to predict irradiating particle and dose rate effects on α' precipitate size, number density and composition.

The paper is organized as follows: Section 2 details the model, including the phase-field formulation, and the models for classical nucleation, radiation-enhanced diffusion, and cascade mixing. This is followed by a general description of simulations predicting the effects of cascade mixing and dose rates on α' precipitation. Section 3 presents the simulation results which are calibrated using experimental data on Fe-12-18% Cr neutron irradiated to 1.8 dpa at ≈ 320°C [9]. In Section 4 we compare the simulation results with recent APT studies of high dose rate heavy-ion and electron irradiations, including temperature effects. We also address the limitations, uncertainties, and potential extensions of the model. Concluding remarks are given in Section 5.

## 2 MODEL

### 2.1 Microscopic phase-field model

We employ the microscopic phase-field model developed originally by Cahn and Hilliard [47] to study the evolution of α' precipitates under irradiation. The model is built upon the Ginzburg-Landau-type free energy functional ($F$) using the Cahn-Hilliard description of chemical non-uniformity:

$$F = \int \left[ f_{ch}(c_{Cr}) + \kappa (\nabla c_{Cr})^2 \right] dV \quad (1)$$

where $c_{Cr}$ is the atomic fraction of Cr (a conserved order parameter characterizing the chemical non-uniformity), $f_{ch}$ is the free energy density (e.g., in units of J m$^{-3}$), and $\kappa$ is the gradient energy coefficient which can be approximated as a function of interatomic distance $\lambda_a$ and regular solution parameter $\Omega$ by $\Omega\lambda_a/2$. Due to the small lattice mismatch between bcc-α (Fe-rich) and bcc-α'(Cr-rich) lattices (≈3%), we ignore the effect of elastic strain energy on α' precipitation in Fe-Cr alloys. The free energy density was obtained from the CALculation of PHAse Diagrams (CALPHAD) database of Fe-Cr system [48] with optimized thermodynamic parameters. Note that there are several updated work on Fe-Cr thermodynamic databases, but here we chose the one assessed by Andersson and Sundman [48]. This database is chosen as it is compatible with the only available Fe-Cr mobility database [49], which was optimized with the diffusion potential given in the work of Andersson and Sundman [48]. Additionally, this assessment [48] is the most commonly-used one for multicomponent commercial alloys. Parameters of the CALPHAD database and Gibbs energy formulations are summarized in the Supplementary Information (SI) Section S1.

The time-evolution of the conserved order parameter $c_{Cr}$ is determined by the Cahn-Hilliard-type equation with a forced mixing term:

$$\frac{\partial c_{Cr}}{\partial t} = V_m^2 \nabla \cdot \left( M \nabla \left( \frac{\delta F}{\delta c_{Cr}} \right) \right) + \left. \frac{\partial c_{Cr}}{\partial t} \right|_{mixing} \quad (2)$$

where $\delta F/\delta c_{Cr}$ is the functional derivative of $F$ with respect to $c_{Cr}$, $V_m$ is the molar volume and $M$ is the composition-dependent chemical mobility that can be derived from the DICTRA database [49]. The magnitude of $M$ will be scaled by the radiation-induced increase of the vacancy concentration as described in Section 2.3. The first term in the equation accounts for thermal diffusion enhanced by irradiation and driven by the gradient of diffusion potential ($\delta F/\delta c$). The second term corresponds to the cascade mixing contribution of damage cascades which will be detailed in Section 2.4. Note that Eq. (2) without radiation effects has been applied to study spinodal decomposition in the Fe-45Cr alloy at 500°C by Xiong *et al.* [50]. Our approach is generally consistent with ref. [50], including the same equation, mobility database [49], and gradient energy coefficient [47] (although with values adjusted to our temperatures). However, we utilize the thermodynamic database assessed by Andersson and Sundman [48], whereas Xiong *et al.* used a self-developed database [50].

*2.2 Nucleation model*

We include nucleation processes in the phase-field model by utilizing a statistically implemented, explicit classical homogeneous nucleation model [34,35]. In view of the stochastic nature of nucleation events, conventional phase-field modeling of a nucleating system is challenging because the evolution of Cr composition is determined by energy minimization and cannot overcome a nucleation energy barrier in the basic formalism. Adding Langevin noise terms to produce necessary fluctuations in chemical or structural non-uniformities is conventionally used to treat nucleation; however, this approach may require artificially large amplitude of fluctuations to induce nucleation events, in the case of high free energy barriers. This difficulty can be avoided by explicitly introducing critical nuclei in the computational supercell based on a Poisson seeding algorithm. The seeding rate is taken as the locally calculated steady-state nucleation rate $J$ characterizing the number of transformed nuclei per phase-field cell of untransformed matrix per unit time, which is given from classical nucleation theory as [51]

$$J = ZN\beta^* \exp\left(-\frac{\Delta G^*}{kT}\right) \tag{3}$$

$Z$ is the Zeldovich factor and can be approximated as $\left(\Delta G^{\text{nucl}}\right)^2 \big/ 8\pi\sqrt{kT}V_a\sigma^{3/2}$, $N$ is the number of available nucleation site per unit phase-field cell, $\beta^*$ is the frequency factor obtained as $4\pi r^* D_{Cr}/V_a$, $k$ is the Boltzmann constant, and $\Delta G^*$ is the local nucleation barrier given by $\Delta G^* = 16\pi\sigma^3/3(\Delta G_{nucl})^2$, where $D_{Cr}$ is the diffusion coefficient of Cr atoms [49], $\sigma$ is the interfacial energy, $V_a$ is the atomic volume, and $r^*$ is the critical radius $2\sigma/\Delta G^{nucl}$. The $\Delta G^{nucl}$ term is the bulk free energy difference between the transformed phase (precipitate) and untransformed phase, which can be calculated directly from the Gibbs energy.

Note that in this model we simulate the nucleation of α' in α matrix without allowing the reverse nucleation direction to occur, i.e., α in α'. Thermodynamically it is possible for an α particle nucleate in the α' matrix if the free energy change of nucleation is strongly negative (e.g., for α' having an Fe supersaturation). While such a situation is not expected under thermal nucleation and growth from an Fe-rich host, under irradiation and significant cascade mixing we do in fact predict a reduced Cr content in the α' precipitate (i.e., Fe supersaturation). In fact, if one just applied classical nucleation theory, α would precipitate rapidly from our Fe supersaturated α' particles. However, we believe that applying classical nucleation theory to α in the α' precipitates is an incorrect approach for the following reason. To precipitate α in an α' particle the system must lower its free energy. This would happen by the Fe-rich a forming inside the α' and enriching the α' in Cr. However, the cascade mixing removes Cr from the α' faster than Cr can accumulate in it, which is why the α' is below its expected Cr composition under normal equilibrium conditions. Therefore, the α cannot form since the α' cannot enrich. Another way of saying this is that under the steady state conditions of cascade mixing the solubility of Fe in the nanoscale α' has effectively been moved to a higher value by cascade mixing, and there is therefore no stable driving force for a formation in α', even when the α' is supersatured in Fe compared to equilibrium conditions. We therefore consider our approximation of excluding nucleation in the direction from α' to α to be appropriate for our simulation conditions. This model is therefore also appropriate for modeling thermal evolution of Fe-Cr provided one starts from a Cr supersaturation, but not appropriate for modeling thermal evolution when starting from an Fe supersaturation in a Cr-rich alloy.

The Poisson seeding algorithm is numerically implemented by calculating the probability of a phase-field cell being transformed into a nucleus in a given time period. Considering that the transformation probability per available nucleation site is $J/N\beta^*$ in one time step of $1/\beta^*$ and assuming that any independent nucleation event of an atomic site can transform the phase-field cell in which the site is located, the transformation probability $p$ of a phase-field cell over a time step $\Delta t$ is given by

$$p = 1 - \left(1 - J/N\beta^*\right)^{N\beta^*\Delta t} \tag{4}$$

where $J/N\beta^* < 1$ was ensured for any Cr composition with negative free energy change of α' nucleation. To determine numerically whether the phase-field cell transforms or not in each time step, a random number is generated for each untransformed cell and the transformation takes place only when the number is smaller than $p$. Note that if the size of the phase-field cell is too large, or the nucleation events occur below the numerically resolvable length scale, the early stages of nucleation and growth may not be captured by this method. We avoid this limitation by choosing the numerical grid size to be the nearest neighbor distance of the bcc-α lattice (~0.249 nm), so only a single nucleation site is available in one phase-field cell. This choice of the grid size ensures that added nuclei occupy at least 4 to 5 grid points in the phase field supercell in any dimensional direction, which is essential for stability of numerical nuclei. As the typical value of the critical radius for α' in our simulations is ~0.5 nm in radius, the choice of ~0.249 nm grid spacing allows 5 grid points along each direction along each direction within a critical nucleus. Concentration conservation was ensured by numerically producing a depletion layer around nuclei following the treatments by Simmons *et al.* [34].

There are multiple approximations associated with our classical nucleation theory approach. First,

such a nucleation model is not appropriate in the spinodal regime. However, we are not in the spinodal regime so the approach is modeling the dominant physics. There is some ambiguity in nucleation mechanisms as one approaches the spinodal barrier [52–54] but as we solidly in the classical nucleation region for our focus of 15% Cr (the spinodal is not until ~25% Cr at 300°C) the classical nucleation approach is appropriate. Even outside the spinodal, for concentrated alloys like our Fe-Cr compositions of interest, the classical nucleation theory may not be the most rigorously correct way to implement nucleation in the phase-field model. Optimization methods could be used to find and follow minimum energy paths to determine the critical nucleus size and composition profile [52,53,55], but the techniques would add significantly to the model complexity and are outside the scope of this work. Such approaches require additional numerical treatments like the nudged elastic band method [55] searching for the minimum energy pathway of nucleation or the saddle-point nucleus. This approach can be treated based on the framework of non-classical nucleation theory of Cahn and Hilliard [56]. Additionally, according to non-classical nucleation theory, early nucleation of clusters with smaller compositional difference with the matrix could be faster than the equilibrium nuclei due to the smaller interfacial energy. Fully correct models should therefore have size and composition dependent interfacial energy. Furthermore, in our classical nucleation theory treatment the seeding nuclei are assumed to have a fixed composition set by thermodynamic equilibrium, and this assumption may induce errors due to the overestimation of the Cr concentration of the critical nuclei. Finally, we note that the classical nucleation theory assumes microstates are appearing consistent with Maxwell-Botlzmann statistics, but under irradiation and cascade mixing it is expected that the microstate distribution will be altered. It is unfortunately difficult to assess the exact scale of errors introduced by treating all these effects with classical nucleation theory. However, as we fit to experimental data and reproduce the observed particle size distributions fairly accurately and with physically reasonable values for all our fitting parameters, it is reasonable to assume our nucleation model is adequate for the present application.

*2.3 Radiation-enhanced diffusion (RED)*

We apply the radiation-enhanced diffusion (RED) model developed by Odette *et al.* [57] to calculate the vacancy concentration under irradiation, $X_V^r$, and scale thermal diffusion coefficients with vacancy supersaturation. The method has been utilized to study radiation-induced precipitation in F/M steels [58]. At steady state, when defect production is balanced by annihilation at sinks as well as recombination in the matrix, particularly at solute trapped vacancies. The model not only considers the vacancy annihilation at sinks and ordinary matrix recombination [59], but also includes the effect of dpa-rate-dependent solute vacancy trapping that enhances vacancy recombination with self-interstitial atoms (SIA), thereby reducing $X_V^r$ [57,60]. SI Section S2 provides detailed descriptions of the RED model that was utilized in this study. Note that flux-dependent unstable matrix defects are considered in this work as an additional sink that reduces $X_V^r$. A steady-state condition of the rate theory equations, as given by Eq. (S10) and Eq. (S11), is assumed in the model, so there is not explicit tracking of the evolution of defect clusters.

The phase field model application in this work requires the vacancy diffusion coefficient, or at least the vacancy migration energy, over almost the entire composition range of Cr to be able to treat the dilute Cr in the host depleted Fe host, the highly concentrated Cr in the α' precipitate, and transition regions in between. Experimental data of composition-dependent diffusion coefficient or vacancy migration energy is not presently available for Fe-Cr alloys and it is challenging to evaluate either by experiments or atomistic simulations [61]. Vacancy diffusion properties in concentrated Fe-Cr alloys were evaluated recently by atomistic studies of radiation-induced segregation (RIS) of Cr [62,63] and decomposition kinetics including the effect of magnetism [64], from which the vacancy migration energy as a function of alloy compositions can be derived explicitly. However, these models were not developed to provide accurate diffusion coefficients for highly Cr-rich concentrations and low temperatures (~300 °C). Therefore, in this study we make a simple and consistent assumption that the vacancy migration energy follows a linear relationship with respect to the Cr content, and fit a linear function to the migration energies of dilute Cr in Fe and self-diffusion in pure Cr. SI Section S2 summarizes this fit and other parameters used in the RED calculation.

*2.4 Cascade mixing model*

Here the continuum atom relocation model developed by Enrique and Bellon is used to treat cascade mixing [41]. The Enrique-Bellon model is a more general approach than the early re-solution model by Nelson *et al*. [24] and the model of effective temperature for cascade mixing derived by Martin [26]. The former assumed that the relocation distance of mixing is arbitrarily large and considered only the transport of atoms from the matrix to precipitates but not vice versa, while the latter assumed that the forced mixing takes place between nearest-neighbor atoms. Note that Frost and Russell considered the recoils of atoms displaced in random directions into the matrix within a sphere of a finite range [27], but this model did not account for the relocation of atoms from the matrix to precipitates. Although some reasonable results can be obtained by following both approaches, these assumptions are not consistent with the general understanding of the finite-range atomic replacements in displacement cascades. The mixing was found to extend beyond the nearest-neighbor distance and it can be approximated well by a distribution function with a average relocation distance $R$ with a magnitude close to the nearest-neighbor distance [44]. Enrique and Bellon characterized the change of local composition caused by the forced cascade mixing as [41]

$$\left.\frac{\partial c_{Cr}}{\partial t}\right|_{mixing} = -\Gamma\left(c_{Cr} - \langle c_{Cr}\rangle_R\right) \tag{5}$$

Here $\Gamma$ is the relocation frequency that is related to the dpa rate ($\phi_{dpa}$) by $b\phi_{dpa}$, where $b$ is a parameter characterizing the number of replacement per atomic displacement, $\langle c \rangle_R$ is the nonlocal average of the concentration weighted by a normalized function (integral is over the entire simulation cell) [41]:

$$\langle c_{Cr} \rangle_R = \int \omega_R(\mathbf{r}-\mathbf{r}')c_{Cr}(\mathbf{r}')d\mathbf{r}' \tag{6}$$

and $\omega_R$ is the normalized function characterizing the distribution of atomic relocations ($R$), which can be described by a Gaussian distribution as [45]

$$\omega_R(\mathbf{r}-\mathbf{r}') = \left(\frac{3}{2\pi R^2}\right)^{3/2} \exp\left(-\frac{3|\mathbf{r}-\mathbf{r}'|^2}{2R^2}\right) \tag{7}$$

Here, $\mathbf{r}$ and $R$ respectively correspond to the spatial location and average jump distance during relocation events. Note the model assumes one-dimensional radial diffusion-mixing symmetry. In principle, both $R$ and $b$ depend on the characteristics of the cascades and thus also on the type of energetic particle. $R$ is typically approximated as the nearest interatomic distance and the value of $b$ ranges from ~1 for electrons to ~30-100 for neutrons and heavy ions [65–68]. Considering that many atoms change positions without creating Frenkel pairs during the cool-down of cascades, the number of replacement is expected to be larger than displacements approximated by the Norgett–Robinson–Torrens model [69] particularly when the cascade is dense and large. In the simulation we use the nearest neighbor distance of the bcc-α lattice ($2.49 \times 10^{-10}$ m) as $R$ and assume that $b = 1$ for electron irradiations and $b = 50$ for neutron/heavy-ion irradiations. A more accurate evaluation of replacement number is desirable and could be determined using molecular dynamics simulations of displacement cascades such as in Ref. [70], but such a study is beyond the scope of this work. However, the results of this study are not highly sensitive to the exact value of $b$, with values for neutrons and heavy ions ranging from 30-100 giving qualitatively similar results, changing the onset of major dose rate effects of less than an order of magnitude from values obtained for $b = 50$.

*2.5 Description of the simulations*

The hybrid microscopic phase-field model with explicit nucleation and forced mixing of cascade was first employed to simulate α' precipitation in Fe-Cr alloys with Cr contents from 12 to 18% during neutron irradiation at 320 °C and at a dose rate of $3.4\times10^{-7}$ dpa/s. The simulation results were compared with the APT analysis of Fe-Cr model alloys under neutron irradiation [9]. By adjusting the interfacial energy of α', vacancy cluster recombination radius and vacancy cluster annealing time within the "reasonable range" of values (by "reasonable range" we mean consistent with previous results and known physics of the relevant length, time, and energy scales, and these consistencies are discuss in detail for the interfacial energy in Section 3.1 and for the recombination radius and vacancy cluster annealing time in SI Section S2 ), we find the optimal fitting value for each parameter to obtain number density, mean radius, and volume fraction of α' precipitates consistent with experiments.

We then employed the same method to explore the effects of flux and cascade mixing on α' precipitation at ~300 °C by changing the dpa rate from the magnitude of typical neutron to typical ion irradiation (higher than $10^{-5}$ dpa/s), exploring a range from $10^{-8}$ to $10^{-3}$ dpa/s. The simulation of the flux effect focused on Fe-15Cr to compare with recent APT studies of heavy-ion irradiated Fe-15Cr alloys. Precipitation of α' in Fe-15Cr under electron irradiation with a weak cascade mixing effect is studied and compared with that under neutron or heavy-ion irradiation. The effect of temperature on α' precipitation under heavy-ion irradiation is also considered.

The primary differences between heavy-ion and electron irradiations are the damage efficiency ($\xi$) and number of atomic replacements per displacement (*b*) in a displacement cascade, both of which are strongly associated with damage morphologies. The former characterizes the net generation of migrating point defects, which enhances kinetics of phase separation, while the latter characterizes the level of forced atomic mixing in displacement cascades, which promotes random solid solution. Under neutron or heavy-ion irradiation, the damage efficiency is similar to that of neutron irradiation which is known to be 0.3–0.4 [71]. Due to the low recoil energy of electrons and low levels of both cascade production under electron irradiation, the damage efficiency is approximately 1 [71,72]. Here we use $\xi = 0.33$ and $b = 50$ for the simulation of neutron or heavy-ion irradiation with strong displacement cascade, and $\xi = 1$ and $b = 1$ for the simulation of weak cascades produced by electron irradiation.

Note that according to the Cahn-Hilliard theory, the interfacial energy is a function of the free energy integral across the interface or phase boundary under local equilibrium, and therefore should be a function of temperature, as given by the following equation [47]:

$$\sigma = 2\int_{c_\alpha}^{c_{\alpha'}} \sqrt{\kappa \Delta f}\, dc_{Cr} \tag{8}$$

Here $c_\alpha$ and $c_{\alpha'}$ are the equilibrium concentrations of α and α' across the miscibility gap and $\Delta f$ is the free energy that is referred by the equilibrium (α+α') mixture as a standard state. Since in this study the interfacial energy is treated as a fitting parameter at 320 °C, its variation with temperature is determined by scaling with Eq. (8), to give a $\sigma(T)$ that is proportional to the integral of $\sqrt{\Delta f}$, which is temperature dependent.

The 3D computational supercell was discretized into a system of 128×128×128 grids and periodic boundary conditions were employed along each direction. The mesh spacing was set to the nearest-neighbor distance of the bcc-α lattice (2.49×10$^{-10}$ m) and the length of the computational supercell was 31.8 nm. The cell size ensured the minimal resolvable number density of precipitate to be at least $10^{27}/31.8^3 = 3.1\times10^{22}$ m$^{-3}$. Eq. (2) is solved by a semi-implicit finite difference scheme and Eq. (5) is calculated in reciprocal space which gives an accurate numerical solution of the cascade mixing term. Note that to ensure numerical stability of the nucleation model, a small time step was implemented throughout the simulations, typically about ~50 s for neutron irradiation at 300 °C. The reduction of the grid size to half of the current value slightly affects the initial nucleation density by less than an order of magnitude and its effect on the diffusion-controlled microstructures and precipitate size is small. To provide a sense of the speed of the calculations, it took approximately 200 h to run a neutron-level irradiation dpa rate of $3.4\times10^{-7}$ dpa/s to 2 dpa at 320°C in OpenMP parallel codes using 12-16 CPUs, each with approximately 60 GFLOPS performance on SPEC FP2006. We note that very little effort was made to optimize the code performance and therefore increased performance is likely possible.

Table 1 lists the parameters use in the phase-field simulations. In assessing precipitate properties for comparison to the APT studies in ref. [9] only the computational grid elements with local composition more than ~80% Cr are identified as α' and volume and size are calculated based on the 80% Cr isosurface. The 80% Cr selection ensures that the plateaus of Cr-rich regions are recognized as α' particles, as displayed in SI Figure S6 showing the composition profile (Figure S6(b)) along a single line in 3D microstructure (Figure S6(a)). It is also worth noting that the profile in Figure S6(b) is quite flat within the α' particle. This result demonstrates that, within this model, even for an α' with significantly reduced Cr composition from ballistic mixing, the average composition is quite representative of the particle Cr content throughout the precipitate. Volume fraction is determined by the formula $N\frac{4}{3}\pi\langle r\rangle^3$, where $N$ is the number of precipitates and $\langle r\rangle$ is their mean radius.

## 3 SIMULATION RESULTS

### 3.1 α' precipitation in neutron-irradiated Fe-Cr alloys at 320 °C

Figure 1 shows the phase-field simulation results for α' precipitation in Fe with Cr contents from 12 to 18% under neutron irradiation at a dpa rate of $3.4\times10^{-7}$ dpa/s to 1.8 dpa at 320 °C. Note that the temperature in the Advanced Test Reactor at Idaho National Laboratory mentioned by Bachhav *et al*. [9,16] was recently corrected from 290 °C to 320 °C. An interfacial energy of 0.090 J m$^{-2}$ was calibrated by fitting the number density, size, and volume fraction of α' precipitates that are consistent with the APT data, as shown in Figure 2 [9]. Note that the optimized value is lower than many other previous values but within the range reported in previous studies based on various modeling methods, including the atomistic simulations, which gave 0.08−0.37 J m$^{-2}$ [74–77], and cluster dynamic semi-empirically fitting with experimental data at 300°C, which gave a much lower value of 0.028 J m$^{-2}$ [78]. The predicted number density of the α' precipitates are about 30-45 % higher than observed by APT. The number density increased significantly from $1.3 \times 10^{24}$ m$^{-3}$ in Fe-12Cr to $7.1 \times 10^{24}$ m$^{-3}$ in Fe-18Cr, reflecting the larger chemical driving force leading to higher nucleation rates at higher solute supersaturation. The corresponding mean predicted α' radius is in excellent agreement with the APT, decreasing with increasing Cr. The volume fractions are also in very good agreement but it should be noted that they are calculated by the method described in Section 2.5. This approach is different than that use for the experimental data by Bachhav *et al*. [9], who estimated the volume fraction based on the number of detected Cr atoms contained in the Cr precipitates. This discrepancy is difficult to correct given the different nature of the experimental and simulation data but is minor and expected to have no impact on the quanlitative conclusions of this work.

Figure 3(a)-(c) shows the evolution of mean radius, number density, and volume fraction of α' as a function of dose up to 2 dpa, again comparing to the data from Bachhav *et al*. [9] (the same differences in volume fraction analysis approaches as discussed above with respect to Figure 2 also occur here). These results show different growth kinetics of α' in Fe-12Cr from the other higher-content alloys. For Fe-15Cr and Fe-18Cr, α' precipitation reaches the coarsening stage at less than 0.1 dpa. This coarsening behavior is similar to the recent KMC study [32] showing that coarsening of α' occurred prior to ~$10^5$ s or 0.034 dpa in Fe-15Cr and Fe-18Cr. The coarsening stage can be identified by the continuous decrease in number density and increase in mean radius and volume fraction in Fe-15Cr and Fe-18Cr. However for Fe-12Cr, the coarsening starts after 0.5-0.8 dpa, which is much later than Fe-15Cr and Fe-18Cr. The difference is caused by the lower nucleation rate in Fe-12Cr, which leads to correspondingly lower number densities and thus larger distances between precipitates. This, in turn, requires longer

diffusion times and higher doses during self-similar coarsening. The Fe-12Cr α' radius also grows much faster than in Fe-15Cr and Fe-18Cr above 0.8 dpa, as shown in Figure 3(a). This is due the fact that the Cr is partitioned to far fewer precipitates. Considering that the oversaturated solute atoms in low Cr matrix such as Fe-12Cr can be absorbed only by a small number of nucleated precipitates through diffusion, the rate of particle size growth is faster compared to high Cr alloys with higher nucleation rate and number density of α' precipitates. The time (or dose) evolutions of full α' particle size distribution (rather than just mean radius) in the neutron-irradiated Fe-Cr alloys are shown in SI Section S3.

As noted above, the simulations slightly overestimate the number density of α'. Further, in the case of the Fe-9Cr alloy at 320 °C, the reduced Cr chemical potential and corresponding nucleation rates are too low to form any α' precipitates in the 31.8-nm simulation cube by 2.0 dpa. This result is not consistent with the experimental study showing the formation of low-density α' in Fe-9Cr under the same neutron irradiation condition [9]. The experimental number density and mean radius were reported as $8.5\times10^{22}$ m$^{-3}$ and 2.4 nm, respectively. This number density is equivalent to only ≈ 3 precipitates in the simulation volume, and this low value may account for their absence during the simulation.

The depleted Cr composition in the matrix is 8.5−9.2% and peaks at ≈ 8.7% at 1.8 dpa in all three Fe-Cr alloys. The matrix composition is slightly lower than that measured by APT at 8.9−10%. The α' Cr concentration is between 80-90% peaking at ≈ 84% in plots of frequency versus local grid values (see Section 4.2), independent of the initial matrix Cr. This composition range is consistent with that measured by APT at 82−90% [9]. Both the simulated and experimental α' Cr compositions are lower than the nominal thermodynamic equilibrium value at 320 °C, which is predicted to be 97.3% by the standard CALPHAD database [48]. Similar lower α' precipitates compositions have been reported in several previous studies of Fe-Cr alloys under neutron irradiation [8,9,79].

The discrepancies described above are relatively minor and can come from many sources, like the uncertainties in CALPHAD and DICTRA databases, with parameterizations that were optimized based on experiments at higher temperatures [48,49]. The uncertainties are discussed in Section 4.5. The discrepancies in $N$ are also not surprising given the extreme sensitivity of nucleation rates to interfacial energy and the free energy evaluations of the α and α' phases.

*3.2 α' precipitation in Fe-Cr during irradiation at 300 °C with high energy primary recoil atom displacement cascades*

To explore the reasons for the observed differences in α' precipitation between neutron and heavy-ion irradiation, we applied the calibrated phase-field model for dose rates from $10^{-8}$ to $10^{-3}$ dpa/s. For simplicity we assumed the same number of replacements per displacement for heavy ions and neutrons. The simulations were performed at 300 °C for a Fe-15Cr alloy to facilitate direct comparisons with experiment [19]. Figure 4(a)-(f) shows the phase-field simulated microstructures in the Fe-15Cr alloy for dose rates from $10^{-8}$ to $10^{-3}$ dpa/s at 10 dpa, with some key quantitative values in Table 2, and the evolutions of number density, mean radius, volume fraction, and precipitate and host Cr content vs. dpa are shown in SI Section S4 (Figure S3). Increasing the dose rate clearly decreases the amount, and affects the nature, of phase separation. At dose rates characteristic of neutron irradiations ($10^{-8}$-$10^{-6}$ dpa/s) increasing dose rate primarily manifests through approximately constant number density, decreasing radius and volume fraction, and decreasing α' Cr composition. The trends in number density, size, and volume fraction are to be expected from the flux effect on RED, which accelerates the nucleation and reduces average size for high-flux irradiation at a given dose or dpa, although the cascade mixing may also be playing some role in these changes. The detailed description of the flux effect is given in Section 4.3. The decreasing Cr composition in α' at increasing dose rates is caused by more frequent cascade mixing replacement events, forcing the exchange of atoms from the matrix to α' particles and vice versa (as described by the forced mixing term in Eq. (5)). Changes in the precipitation behavior are expected to be due to changes in the RED coefficient $D^*_{RED}$, the ballistic relocation frequency $\Gamma$, and in particular their ratio $\gamma = \Gamma / D^*_{RED}$ [41], which largely controls the precipitate stability, as this represents the balance between destabilizing cascade mixing and kinetics that allows precipitate formation. The changing precipitate composition and volume fraction with increasing flux should therefore be understood as both due to increased cascade mixing and increased $\gamma$. This ratio changes because $\Gamma$ increases linearly with flux, while $D^*_{RED}$ will increase less than linearly due to recombination, reaching a square root dependence on flux in the recombination dominated limit [57] (see Section 4.3 for more information on these trends). The transition from a phase separating to solid solution domain with increasing $\Gamma$ is consistent with the phenomenological trends identified by similar models [41,80].

As can be seen in Figure 4, and more extensively in Table 2 and SI Figure S3, the trends in radius, number density, and volume fraction change dramatically at about $10^{-5}$ dpa/s. At this dpa rate some precipitates still form and grow, but the nucleation and growth are dramatically slower vs. dpa. Then by $10^{-4}$ dpa/s no stable precipitates are observed at any of the dpa rate values studied. Thus the model predicts a transition to a dpa rate condition where there are no precipitates found to occur somewhere between $10^{-5}$ and $10^{-4}$ dpa/s. This transition is due to the increasing role of mixing relative to diffusion and the increasing value of $\gamma$, which reaches a value so large that stable precipitates cannot nucleate and grow. Overall, at $\leq 10^{-6}$ dpa/s the effects on $r$, $N$ and $f$ are primarily due to the effects of dose rate on $D^*_{RED}$, rather than mixing effects per se. However, at higher dose rates $\geq 10^{-5}$ dpa/s mixing effects are dominant.

While diffuse Cr clustering is observed at the intermediate rate of $10^{-5}$ dpa/s, as shown in Figure 4(d) only ~5 α'-type precipitates form in the 31.8-nm cube at 10 dpa. $N \approx 1.6 \times 10^{23}$ m$^{-3}$ represents an order of magnitude reduction compared to lower dose rates. The α' composition of ~78.8% is also lower than found at lower neutron relevant dpa rates. Note that there are other small Cr-rich clusters with mean radius less than 0.5 nm in the simulations. These clusters continuously form and dissolve due to the cascade mixing. The concurrent radiation-induced dissolution and re-precipitation at $10^{-5}$ dpa/s is illustrated in Figure 5 (a)-(c) at 2, 5, and 10 dpa. Most of the newly formed clusters are kinetically unstable and only a few are able to grow larger than 0.5 nm. At higher dose rates of $10^{-4}$ and $10^{-3}$ dpa/s α' cannot form a stable precipitate at all, as seen in Figure 4(e) and (f). However, dilute Cr clusters do form, but all are later re-dissolved.

*3.3 α' precipitation in Fe-Cr at 300 °C with Frenkel pair defect production simulating electron irradiations*

In this section, we explore the effect of displacement damage, primarily in the form of one or a few vacancy-self interstitial atom Frenkel pairs, characteristic of electron irradiations. The simulations use the same calibrated phase-field α' precipitation thermo-kinetic parameters, but with a few key changes (see Section 2.4): a much lower mixing rate ($b$ = 1/dpa vs. $b$ = 50/dpa for heavy ions, a higher efficiency ($\xi$ =1 vs. $\xi$ =0.33 for heavy ions), and no vacancy clusters in the RED model. Figure 6(a)-(f) shows the simulation results for the Fe-15Cr alloy irradiated at 300°C to 10 dpa under Frenkel pair irradiation conditions, again from $10^{-8}$ to $10^{-3}$ dpa/s (the evolution of number density, mean radius, volume fraction, and precipitate and host Cr content vs. dpa for these electron irradiations are shown in SI Section S5). The results are summarized in Table 3, which shows that, as in the case of neutron/heavy-ion cascade mixing irradiations, the α' $r$ and $f$ decrease, while $N$ increases, with increasing dose rate. Again, as discussed for neutron/heavy-ion irradiation, these effects can be understood as due to increased $D^*_{RED}$ with increased dpa rate. However, phase separation of α' is predicted at all dose rates from $10^{-8}$ to $10^{-3}$ dpa/s. The result is in contrast to the neutron/heavy-ion irradiation results (Section 3.2) showing much less phase separation for dose rates at and above $10^{-5}$ dpa/s. Also different from the neutron/heavy-ion simulations is that in the case of Frenkel pair irradiations, the α' Cr composition is more than 90%, just slightly below the equilibrium value, at all the dpa rates, and the α' Cr concentration decreases only slightly with increasing dose rate. These simulations are consistent with the accelerated α' precipitation observed experimentally in electron irradiations at a dose rate $3.9 \times 10^{-5}$ dpa/s [22], as discussed in the next section.

These electron irradiation modeling results suggest that the weak cascade mixing in the electron irradiation model has a small effect on α' evolution. To confirm that the mixing plays a small role in the electron irradiation model, we show the evolution of number density, mean radius, volume fraction, and precipitate and host Cr content vs. dpa for these electron irradiations without cascade mixing in SI Section S6 (Figure S5). These simulations are identical to those shown in Figure S4 except that cascade mixing from the electrons has been turned off. The nearly identical results obtained with and without cascade mixing demonstrate that it is playing almost no role in the electron irradiation model, and that the changes with dpa rate are due to changes in the RED (changes in RED are discussed further in Section 4.3). Overall, our results predict that under Frenkel pair electron irradiation, phase separation follows an approximately normal thermodynamic path at a rate that is accelerated by RED, and is little affected by mixing effects even under very high dose rate conditions.

## 4 DISCUSSION

### 4.1 Comparison with experimental data

The phase-field simulations in this study are in generally good agreement with experimental observations of Fe-Cr model alloys irradiated with neutrons, heavy ions and electrons. Under high-flux irradiations that produce displacement cascades (heavy ions and neutrons), the phase-field simulations (Section 3.2) clearly show that cascade mixing inhibits the formation of α' precipitates, and even causes them to dissolve, at high dpa rates that are typical of heavy-ion irradiations. Thus the phase-field model rationalizes the previously unexpected discrepancies between low-flux neutron and high-flux ion irradiations observed in recent experiments, where α' precipitates were not found at high dpa rates [18,19]. Our simulations of electron irradiations that produce only Frenkel pairs, with much less mixing compared to heavy-ion irradiations further support the conclusion that the changes seen in α' for high-flux heavy-ion experiments are due to the strong mixing effects of cascades. The Frenkel pair electron irradiation results show that α' precipitates form, even at very high dpa rates up to $10^{-3}$ dpa/s. Indeed, as shown in SI Figures 3 and 4, and Table 3, the Frenkel pair electron irradiation precipitation behavior is almost identical to simulations with mixing effects completely eliminated. Comparisons with experiment are discussed further below.

The model also predicts that cascade mixing leads to reductions in the Cr content of α' compared to thermodynamic equilibrium compositions, even at dose rates relevant to some neutron irradiations. The predicted α' Cr composition for a 300 °C neutron irradiation at $10^{-7}$ dpa/s is between 80 and 90%, well below the thermodynamic value of 97-98% [48]. Note that for dose rates $\leq 10^{-7}$ dpa/s we believe that the α' compositions and volume fractions at 10 dpa have reached approximately steady-state and will not change with further simulation to higher dpa except through very slow coarsening processes as displayed in SI Figure S3. In contrast, the predicted Cr contents of the α' precipitates that form under Frenkel pair electron irradiations are higher than 90%, even at $10^{-3}$ dpa/s. Thus the simulations demonstrate that that the lower Cr in α' for irradiations by heavy ions is largely caused by the cascade mixing.

These predicted trends are broadly consistent with experiment. For example, a recent APT study of a neutron irradiated Fe-15Cr alloy found α' Cr compositions from 80-90% for precipitates up to ≈ 2 nm and an even higher average at larger sizes [9,81]. As expected electron irradiations of a Fe-15% alloys found ≈ 96% Cr in α' for larger ≈ 3 nm radius precipitates. However, the Cr contents decreased in smaller α' precipitates to slightly less than 50% at ≈ 1 nm. [43]. Other data in literature shows a wide variety of Cr compositions ranging from 40-80 [15] and 55-60% [8,12] for neutron irradiations, which is much less than predicted by the phase-field simulations. However, the APT compositions are strongly dependent of the precipitate size, especially below 1.5 nm [15,81]. The Fe content α' in precipitates less than 1-2 nm in radius is overestimated in APT, primarily due to so-called trajectory aberrations, that result in an interface mixing zone containing both matrix and precipitate atoms [82,83]. It is therefore possible that this discrepancy between the higher simulation and lower APT Cr compositions is due at least in part to experimental artifacts. The variation in α' composition that has been observed in thermal ageing studies has been attributed to changes with time or dose, rather than APT artifacts [84,85]. However, both are related to the precipitate size thus the two effects are difficult to deconvolute from one another. Additional exploration of this issue would be valuable, but further discussion of this complex topic is beyond the scope of this paper.

As illustrated in Figure 4(c) to (e), the phase-field simulations (Section 3.2) indicate a transition between stable α' nucleation and growth at a cascade mixing heavy ion dpa rates of $10^{-6}$ dpa/s and at $10^{-4}$ dpa/s, where a dissolution-dominated microstructure, that is nearly absent any significant Cr clustering, is predicted. The intervening condition at $10^{-5}$ dpa/s is primarily characterized by continuously precipitating and re-dissolving dilute Cr clusters along with a few α' precipitates. These predictions are semi-quantitatively consistent with heavy-ion irradiation experiments at 300 °C showing an absence of stable α' precipitates in a Fe-15Cr alloy at dose rates from $10^{-5}$ to $10^{-4}$ dpa/s [18,19]. Another APT study of a Fe-15Cr alloys reported that 300 °C heavy-ion irradiations at $3\times10^{-5}$ and $6\times10^{-5}$ dpa/s to 0.5 and 0.8 dpa, resulted in a high density (≈ $2\times10^{24}$ - $8\times10^{24}$ m$^{-3}$) of small Cr-enriched clusters [86] but no α' precipitates. The mean radius of these clusters, that contained less than ≈ 50% Cr, was ~1.0 nm. Note that, while the APT estimates of the cluster Cr content are not reliable, the absence of α' is a clear indication of the dominance of cascade mixing. These results are consistent with Figure 5 which shows that the phase-field simulations predict at $10^{-5}$ dpa/s and 300 °C a high density of ~$3\times10^{24}$ m$^{-3}$ Cr-rich clusters, due to concurrent re-precipitation and dissolution. However, most of these features are smaller than 0.6 nm, and far fewer clusters ($\approx 1.6\times10^{23}$ m$^{-3}$) continuously grow as stable α' precipitates. Thus both experiment and the phase-field simulations indicate that in a Fe-15Cr at 300 °C alloy the critical dpa rate resulting in the instability of α' precipitates is slightly more than $\approx 10^{-5}$ dpa/s. Note these critical rates are lower than in most heavy-ion irradiation experiments. The phase-field model results are also consistent with the results of a 300 °C electron irradiation for a Fe-15Cr alloy at $3.9\times10^{-5}$ dpa/s that results in formation of stable α' precipitates containing more than 90% Cr [22]. Note, the RED model parameters were fitted to the neutron irradiation data, and for quantitative modeling of the electron irradiation may require further assessment and optimization.

*4.2 Effect of cascade irradiation on degree of phase separation of α'*

The radiation effect on the degree of phase separation for cascade mixing under heavy-ion irradiations from $10^{-8}$-$10^{-3}$ dpa/s can be partially quantified in terms of the frequency a specified Cr composition in a grid of voxels set by an a ≈ 0.249 nm grid spacing (voxel volume is a$^3$) in intervals of 1%. The histogram peaks at low and high Cr are assumed to represent the compositions in matrix and α', respectively. For dose rates up to $10^{-6}$ dpa/s the curves indicate strong phase separation behavior, which is manifested by the higher Cr composition peaks and a shift of the other peak to lower matrix Cr. In contrast, at dpa rates from $10^{-4}$ to $10^{-3}$ dpa/s the higher Cr peak is completely absent and the lower Cr peak remains near the bulk alloy composition. However, the extended low Cr peak tail indicates limited solute clustering, even at the highest dpa rates studied. A clear transition occurs at $10^{-5}$ dpa/s, with a largely unaffected low Cr peak with an extended tail ending at a small cluster peak at 78.5% Cr, consistent with the changes noted in precipitate behavior in Section 3.2. The extended tail at $10^{-4}$ and $10^{-3}$ dpa/s is due to the concurrent re-precipitation and dissolution.

*4.3 Flux effects on RED and associated precipitate evolution vs. dose*

As discussed above in Section 3.2 and 3.3, flux effects on $D^*_{RED}$ play a critical role in understanding the flux dependence of α' precipitate evolution with dose. Therefore, in this section we further clarify how flux couples to $D^*_{RED}$ and the associated precipitate evolution with dose, not considering any effects of cascade mixing. In the defect recombination limit $D^*_{RED}$ increases roughly with the square root of dpa rate instead of linearly with dpa rate, where the latter scaling would be the case absent recombination mediated dose rate effects. A plot of $D^*_{RED}$ versus dpa rate for a highly physical model is shown in SI Figure S1, and the non-linear scaler of $D^*_{RED}$ vs. dpa rate can be seen throughout the dpa rate range considered. Thus, to produce a given amount of precipitation, which scales with $D^*_{RED} \times t$ (where $t$ is time) the required dpa increases with increasing dpa rate. This effect can be treated in terms of an effective dose, $dpa_e$, which can be related to the actual dpa rate as $dpa_e = dpa[(dpa/s)_r/(dpa/s)]^p$, where $(dpa/s)_r$ is an arbitrary reference dpa rate. The power $p$ depends on the ratio of defects reaching sinks, that control $D^*_{RED}$, versus those that suffer recombination. The dpa rates in this study are in the recombination dominated regime, where $p = 0.5$. In the sink dominated regime $p = 1$, and since the time to reach a certain dose scales as 1/dpa rate and the kinetics of the system scales as dpa rate, essentially the same precipitate evolution vs. dpa are obtained and there is no dose rate effect. Thus a strong flux dependence of precipitate evolution is expected in this work independent of the cascade mixing effects, as clearly seen for our low flux neutron/heavy-ion (discussed in Section 3.2) and for electron (discussed in Section 3.3) irradiation simulations. Note, the dominance of recombination is associated with solute vacancy trapping, that increases the steady-state vacancy concentration, thus the corresponding recombination rate [57].

*4.4 The effect of temperature on α' precipitation under irradiation*

Eq. (2) shows that the effect of displacement cascades on phase separation is determined by the relative magnitudes of thermo-kinetic clustering and cascade mixing contributions. The former reflects free energy driving pressures ($\Delta G$) as well as $D^*_{RED}$. $D^*_{RED}$ scales with the steady-state point defect concentration. Both $\Delta G$ and $D^*_{RED}$ depend strongly on temperature. In contrast, cascade mixing is athermal. Thus changes in temperature modify the effect of flux on phase stability. These modifications may favor or oppose phase separation. For example, increasing temperature from 300 to 400 °C increases the thermal diffusion coefficient of Cr in bcc-α Fe by 3-4 orders of magnitude (which translates into a large increase in $D^*_{RED}$ through the enhanced movement of vacancies), but reduces $X_V^r/X_V^e$ or radiation enhancement by ~2 orders of magnitude as more point defects can be annihilated at sinks avoiding recombination. Higher temperature also reduces $\Delta G$ and the associated phase separation driving force, which reduces the nucleation rate, but in the Cahn-Hilliard theory this is partly compensated by a corresponding decrease of interfacial energy [47]. Thus multiple competing factors will determine the net influence of changes in temperature. In the case of the Fe-15Cr alloy, higher temperature reduces the effect of cascade mixing primarily due to the large increase in Cr diffusion coefficients, but this trend may vary for different systems and temperature ranges.

To explore the effect of temperature on α' precipitation under neutron/heavy-ion irradiation, simulations using the same thermodynamic and diffusion expressions as above but at 400 °C (see SI Section S1 and S2 for relevant expressions to determine values) are shown in Figure 8 (a) and (b) for dose rates of $10^{-5}$ and $10^{-4}$ dpa/s, respectively. Precipitation of α' is predicted in both cases in contrast to the corresponding behavior at 300 °C, shown in Figure 4(e) and (f). At 400 °C increasing the dpa rate from $10^{-5}$ to $10^{-4}$ dpa/s only slightly increases the matrix composition from 11.5% to 12.5% Cr and reduces slightly the α' composition from 84.6 to 82.5% Cr. Note that for Fe-15Cr irradiated at $10^{-4}$ dpa/s, some concurrent re-precipitation and dissolution takes place and results in some small Cr-rich clusters, as seen in Figure 8(b). This suggests that $10^{-4}$ dpa/s is slightly lower than the critical dpa rate that results in complete phase instability of α' at 400 °C. In summary, within specified limits associated with the dominance of diffusional kinetics, increasing the temperature increases the critical dpa rate for the suppression of phase separation. However, the effect of lower $\Delta G$, especially on the nucleation rate as the Cr solvus is approached, must inevitably reverse this trend, as going above the Cr solvus leads to no precipitation.

*4.5 Potential applications, limitations and insights of the model*

The model provides a new computational tool integrating nucleation theory, Cahn-Hilliard-type phase-field model of precipitation and forced cascade mixing in displacement cascades into a single framework. Thus model can readily be applied to other binary or multicomponent systems as long as the databases for thermodynamic and kinetic parameters are available, and a reasonable RED model is available. The microscopic phase-field theory allows the exploration of microstructural features in practical time and length scales of interest to model very complex irradiation effects. Note that the phase-field model could be extended to dislocations and dislocation loops [87] to simulate their effects of radiation-induced segregation and precipitation.

As mentioned briefly in Section 3.1, there are some disagreements between experiments and simulations, which may be due to the reliability of the thermodynamic and kinetic databases. In particular, the DICTRA and CALPHAD databases [48,49] used in the present study were assessed at temperatures higher than 450 °C, and therefore may not provide accurate descriptions of thermodynamic and diffusion properties at ~300 °C. *Ab initio* calculations [88–91] show a negative mixing enthalpy at low concentrations of Cr which results in a higher solubility limit of Cr in the α phase. Specifically, Bonny *et al*. suggested that the phase boundary in the Fe-rich region shifted to higher ≈ 7.8% Cr at 300°C [92], and further re-parameterized an updated CALPHAD database [93]. Nevertheless, the suggested phase boundary in the Fe-rich region was partly based on the experiments under irradiation, which could potentially modify the solvus line. This issue was critically addressed by Xiong *et al.* in their comprehensive review of the Fe-Cr system [94]. Xiong *et al.* also proposed an improved thermodynamic assessment, predicting that the solubility of Cr in bcc-α of about 9.6% at 300 °C [95]. Notably, the first principal results are very consistent with recent APT studies of Bachhav *et al.* [9] and Reese *et al.* [81,96] who observed precipitation in 9 to 18Cr alloys over a range of temperatures, including the neutron irradiations at 320 °C used to calibrate the model, in a dose rate regime minimally affected by cascade mixing. However, the updated thermodynamic parameterization for the Fe-Cr system does not provide adequate kinetics of nucleation and growth of α' in Fe-12Cr, contrary to observation. This is likely due to limitations in the homogeneous nucleation model, which for example, does not treat heterogeneous mechanisms. A more complete model for Fe-Cr α' evolution that incorporates all heterogeneous nucleation mechanisms and the most accurate thermodynamics is beyond the scope of this work, but would not be expected to impact the qualitative conclusions derived here related to flux and cascade mixing effects.

The very recent KMC study by Soisson *et al*. [97] demonstrated that, contrary to their earlier study [32], the cascade mixing effect is able to cause α' dissolution under high-flux irradiation of heavy ions. The ballistic or cascade mixing effect was considered and the general result is in qualitative agreement with our present study. A cluster dynamics model was used to evaluate the evolution of defect clusters and thus the work considered the sink strength as a flux-dependent and dose dependent property [97]. One major discrepancy of the simulation result from our study is that Soisson *et al*. predicted a sudden dissolution of α' after certain amounts of precipitation when the dpa rate of heavy-ion irradiation is higher than a moderate dose rate ($5.2 \times 10^{-5}$ dpa/s). The discrepancy is possibly caused by the time-dependent sink strength treated by Soisson *et al*. [97], whereas our model considered a steady-state condition. It should also be noted their cluster dynamics modeling of defect clusters [97] did not consider the potential overlap or spatial exclusion by other defect clusters which causes the matrix frustration [98] or volume exclusion [99,100] effect. This effect may become important when the defect cluster density is high, particularly under high-flux heavy-ion irradiation. In our study, we assumed the exclusion area is in an approximated radius of ~10 nm, which results in the saturated sink strength of defect clusters to be ~$3\times10^{15}$ m$^{-2}$. The cluster dynamic model in ref. [97] predicted that the sink strength can reach $10^{18}$ m$^{-2}$ under high-flux heavy-ion irradiation ($2.8 \times 10^{-3}$ dpa/s irradiated to $10^4$ s) and still increased further with dose. This sink strength magnitude indicates that the distance between defect clusters is less than 1 nm, which would be altered significantly by including an exclusion effect.

It is useful to briefly consider the relative pros and cons of using a phase field method, as in this work, vs. the KMC approach, e.g., as used by Soisson *et al*. [32,97]. In general phase field approaches can reach much larger length and time scales than KMC as they are not restricted to consider atomic length scales, but due to the atomic scale grid size necessary to capture the small critical nuclei in this work (see Section 2.2) our phase field simulation time and length scales are similar to those from KMC. Furthermore, KMC is a powerful method to simulate microstructural evolution with full atomic resolution and can model stochastic nucleation processes in a more natural and straightforward fashion than phase field. Thus KMC has many advantages over phase field for both insight and quantitative simulation. However, to use KMC an atomistic Hamiltonian must be established. To do this with our thermodynamics and kinetic approach the CALPHAD thermodynamics and kinetics would need to be mapped onto atomic interactions, which has no unique mapping and could yield many possible models. Ab initio methods and cluster expansion could be used to determine these interactions but this would greatly expand the scope of this study and introduce additional uncertainties associated with using ab initio methods. Thus the main advantage of pursuing phase field for the specific study given here vs. KMC is the ease and uniqueness of mapping the CALPHAD thermodynamics and kinetics into the model. Further length and time scale advantages for phase field vs. KMC could likely be realized by careful parallelization and remeshing to a larger grid after the nucleation stage, but such improvements were not necessary and therefore not pursued in the present work. Another mechanism that has been proposed to explain the retardation of α' precipitation during high-flux neutron/heavy-ion irradiation is the effect of injected interstitials and the creation of high-density sinks such as dislocation loops [83]. This study suggested that injected interstitials could increase the recombination rate and create defect sinks to reduce vacancy concentration and RED. Given that injected interstitials are a relatively small fraction of the total produced interstitials, it is unlikely that their presence would directly alter vacancy concentration enough to significant impact RED and associated precipitation. However, they can alter the balance of interstitials and vacancies, which can have a significant impact on microstructural evolution, e.g., suppressing void swelling or leading to enhanced interstitial loop formation. In our model the evolution defect sinks such as vacancy clusters and small dislocation loops that form during neutron/heavy-ion irradiation are not explicitly modeled, but their effects are approximately taken into account in the RED model through unstable matrix features, which promote recombination and reduce RED. Note that the models developed here are most applicable for regions without injected interstitials.

It is interesting to note that, in contrast to Cr-rich α', Cu-rich precipitates readily form in an Fe host under similar 300 °C heavy-ion irradiations at high dose rates [101,102], and the same has been recently shown for Cu-bearing RPV steels [103]. The difference between Cu and Cr precipitation behavior is most likely due to the fact that at 300 °C the supersaturation of Cu in Fe, driving clustering, is much higher than that for Cr, with a calculated energy hump or barrier of $f_{ch}$ across the miscibility gap 4 to 5 times higher [48,104,105]. Thus is expected that much higher flux of heavy-ion dose rate is required to destabilize Cu-rich precipitates. This is also the case for nanoscale oxides in oxide dispersion-strengthened steels which remain unaffected in very high dose rates [106].

# 5   CONCLUSIONS

We have developed an integrated model of precipitation to study the effect of dose rate and the type of primary damage on phase separation of α' in Fe-Cr alloys. The model is based on the Cahn-Hilliard phase-field theory parameterized by CALPHAD and DICTRA databases of the Fe-Cr system. The model also treats two crucial radiation effects: radiation-enhanced diffusion and cascade mixing. Nucleation is treated explicitly by a classical homogeneous model and incorporated in the phase-field simulation by a Poisson seeding algorithm. The model was calibrated to data for Fe-Cr binary model alloys containing 12 to18 at.% Cr that were neutron irradiated at 320 °C to 1.8 dpa.

The simulations show significantly reduced α' formation with an increasing flux for cascade mixing neutron/heavy-ion irradiations along with lower precipitate Cr contents. Most dramatically, at rates $10^{-5}$-$10^{-4}$ dpa/s and above, α' precipitates are predicted not to form. At high dpa rates $\geq 10^{-4}$ dpa/s, characteristic of heavy-ion irradiations, the Fe-15Cr microstructure is dominated by concurrent Cr clustering and cluster dissolution due to cascade mixing. The simulations are very consistent with a variety of experimental trends, and are the first to successfully rationalize the discrepancy between low-flux neutron and high flux heavy-ion irradiations. The simulations also predict that electrons, which produce Frenkel pairs rather than cascades, with little mixing but significant RED, accelerate but do not alter the basic nature of thermodynamically mediated α' precipitation. In this case near equilibrium α' evolves even at very high dpa rate of $10^{-3}$ dpa/s. The absence of mixing effects in this case is also shown by simulations with no mixing that are nearly identical to the Frenkel pair electron irradiation results. These phase-field predictions are also consistent with a recent experimental study of accelerated near equilibrium α' precipitation under electron irradiation. While mixing effects are minimal in Frenkel pair electron irradiations, dose rate influences on RED mediated changes in the α' size, number density, and volume fraction are profound.

Limited simulations of dose rate and mixing effects at a higher temperature of 400°C show that α'

is more stable at high flux than at 300°C. This result is consistent with higher rates of diffusional clustering (recovery) relative to athermal mixing processes, and may also be due to the precipitates having larger sizes. Most significantly the extended phase-field simulations that include realistic treatments of both mixing and RED paint a consistent picture of α' precipitation over a wide range of conditions.

The results demonstrate a critical role for flux effects on α' precipitate in Fe-Cr through flux dependent RED and cascade mixing. While these processes are generally relevant, many factors, including temperate, size, precipitate stability and free energy driving precipitation, and diffusion kinetics, may all influence results dramatically, so each materials system and set of conditions must be considered individually.

**Acknowledgements**

This work at the University of Wisconsin-Madison, University of Michigan, and UC Santa Barbara (UCSB) was initially funded by the US Department of Energy (DOE), Office of Nuclear Energy (NE), Integrated Research Project (IRP) titled "High Fidelity Ion Beam Simulation of High Dose Neutron Irradiation" (DE-NE0000639). Funding of the final calculations and paper writing was provided by the US Department of Energy Office of Nuclear Energy's Materials Research Pathway for the Light Water Reactor Sustainability Program.

**FIGURE CAPTION**

Figure 1. Simulation results showing α' precipitation in neutron-irradiated (3.4×10$^{-7}$ dpa/s) Fe-12-18 at.% Cr at 320 °C. The color shows the composition of Cr at the local grid points in the 31.8-nm computational supercell.

Figure 2. Plots of comparison showing the (a) mean radius, (b) number density, and (c) volume fraction between the phase-field (PF) simulation results and experimental values reported by the APT study [9].

Figure 3. Plots showing the predicted evolution of the (a) mean radius, (b) number density, and (c) volume fraction of α' precipitate in neutron-irradiated (3.4×10$^{-7}$ dpa/s) Fe-12-18 at.% Cr at 320 °C. The volume fraction is evaluated by using the 80 at.% Cr isosurface. The solid symbols show the experiment measurements [9].

Figure 4. Simulation results showing α' precipitation in Fe-15Cr at 300 °C irradiated to 10 dpa at the dpa rate: (a) 10$^{-8}$ dpa/s, (b) 10$^{-7}$ dpa/s, (c) 10$^{-6}$ dpa/s, (d) 10$^{-5}$ dpa/s, (e) 10$^{-4}$ dpa/s, and (f) 10$^{-3}$ dpa/s. The color shows the composition of Cr at the local grid points in the 31.8-nm computational supercell. Simulations are for neutron or ion irradiation conditions and therefore include strong cascade mixing (see Section 2.4).

Figure 5. Simulation results showing the evolution of α' precipitation at 10$^{-5}$ dpa/s at 300 °C after irradiation to (a)2, (b)5, and (c)10 dpa. The purple color shows the isosurface of 75% Cr. Simulations are for neutron or heavy-ion irradiation conditions and therefore include strong cascade mixing (see Section 2.4).

Figure 6. Simulation results showing α' precipitation in Fe-15Cr at 300 °C irradiated to 10 dpa at the dpa rate: (a) 10$^{-8}$ dpa/s, (b) 10$^{-7}$ dpa/s, (c) 10$^{-6}$ dpa/s, (d) 10$^{-5}$ dpa/s, (e) 10$^{-4}$ dpa/s, and (f) 10$^{-3}$ dpa/s. The color shows the composition of Cr at the local grid points in the 31.8-nm computational supercell. Simulations are for electron irradiation conditions with only Frenkel

pair production and therefore include weak cascade mixing (see Section 2.4).

Figure 7. Composition histogram plots showing the Cr concentration in the grid volumes for Fe-15Cr irradiated to 10 dpa at 300 °C (see Figure 4) for dpa rates from $10^{-8}$ to $10^{-3}$ dpa/s. The lines connect points on a histogram with a Cr binning with of 1%.

Figure 8. Simulation results showing α' precipitation in Fe-15Cr at 400 °C irradiated to 10 dpa with strong displacement cascades at the dpa rate: (a) $10^{-5}$ dpa/s and (b) $10^{-4}$ dpa/s. The color shows the composition of Cr at the local grid points in the 31.8-nm computational supercell.

Table 1. Parameters for the phase-field model of α' precipitation under irradiation. The variables marked with a "*" are determined by fitting with the neutron irradiation experiment [9].

| Parameter | Symbol | Value |
|---|---|---|
| Interatomic distance | $\lambda_a$ | $2.49 \times 10^{-10}$ m |
| Grid spacing | $d_{grid}$ | $2.49 \times 10^{-10}$ m |
| *α' interfacial energy | $\sigma$ | 0.090 J m$^{-2}$ (320 °C) |
| *Vacancy cluster recombination radius | $r_c$ | $3.0 \times 10^{-10}$ m [60] |
| Vacancy cluster production cross-section | $\sigma_c$ | $1.5 \times 10^{-29}$ m$^2$ [57] |
| *Vacancy cluster annealing time | $\tau_c$ | $8.1 \times 10^{-12}/e^{(-1.70/kT)}$ [60] |
| Cascade efficiency | $\xi$ | 0.33 (neutron/heavy-ion irradiation) [71]<br>1.0 (electron) [71] |
| Replacements per displacement | $b$ | 50 (neutron/heavy-ion irradiation)<br>1 (electron irradiation) |
| Average relocation distance | $R$ | $2.49 \times 10^{-10}$ m [44] |
| Atomic volume | $V_a$ | $1.18 \times 10^{-29}$ m$^3$ |
| Dislocation sink strength | $S_d$ | $1.0 \times 10^{13}$ m$^{-2}$ |
| Matrix recombination radius | $R_r$ | $5.7 \times 10^{-10}$ m [57] |
| Trap recombination radius | $R_t$ | $5.7 \times 10^{-10}$ m [57] |
| Binding energy for trapped vacancies | $H_b$ | -0.094 eV [73] |

Table 2. The characteristics α' precipitates in Fe-15Cr irradiated at various dose rates ($10^{-8}$-$10^{-3}$ dpa/s) at 10 dpa at 300 °C. The values marked with a "*" are determined by the 75% Cr isosurface due to the lower α'composition (< 80% Cr). Note that the clusters with mean radius less than 0.5 nm were excluded in the calculations. Simulations are for neutron or heavy-ion irradiation conditions and therefore include strong cascade mixing (see Section 2.4).

| dpa rate (dpa/s) | Mean radius (nm) | Number density (m$^{-3}$) | Volume fraction (%) | Cr at.% in α' | Cr at.% in α |
|---|---|---|---|---|---|
| $10^{-8}$ | 2.5 | $1.1 \times 10^{24}$ | 7.6 | 85.2 | 6.2 |
| $10^{-7}$ | 2.2 | $1.8 \times 10^{24}$ | 6.3 | 83.3 | 7.0 |
| $10^{-6}$ | 2.1 | $1.5 \times 10^{24}$ | 5.5 | 81.4 | 8.1 |
| $10^{-5}$ | *1.3 | *$1.6 \times 10^{23}$ | *0.17 | 78.5 | 14.4 |
| $10^{-4}$ | - | - | - | - | 14.7 |
| $10^{-3}$ | - | - | - | - | 14.9 |

Table 3. The α' $r$, $N$ and $f$ in Fe-15Cr irradiated to 10 dpa at 300 °C at dose rates from $10^{-8}$ to $10^{-3}$ dpa/s. Simulations are for electron irradiation conditions with only Frenkel pair production and therefore include weak cascade mixing (see Section 2.4). The values in brackets are the results for electron irradiation without mixing.

| dpa rate (dpa/s) | Mean radius (nm) | Number density (m$^{-3}$) | Volume fraction (%) | Cr at.% in α' | Cr at.% in α |
|---|---|---|---|---|---|
| $10^{-8}$ | 2.2(2.3) | 1.5×10$^{24}$(1.5×10$^{24}$) | 7.0(7.1) | 95.7(97.4) | 5.8(5.7) |
| $10^{-7}$ | 1.9(1.9) | 2.3×10$^{24}$(2.2×10$^{24}$) | 6.5(6.6) | 95.5(97.2) | 6.1(5.9) |
| $10^{-6}$ | 1.6(1.6) | 3.4×10$^{24}$(3.3×10$^{24}$) | 5.9(6.1) | 95.0(97.0) | 6.3(6.2) |
| $10^{-5}$ | 1.4(1.4) | 4.1×10$^{24}$(4.3×10$^{24}$) | 5.5(5.5) | 94.7(96.8) | 6.6(6.4) |
| $10^{-4}$ | 1.3(1.3) | 4.5×10$^{24}$(4.7×10$^{24}$) | 4.4(4.5) | 93.7(95.9) | 7.8(7.7) |
| $10^{-3}$ | 0.8(0.8) | 5.0×10$^{24}$(5.3×10$^{24}$) | 1.2(1.3) | 92.5(95.0) | 13.4(12.8) |

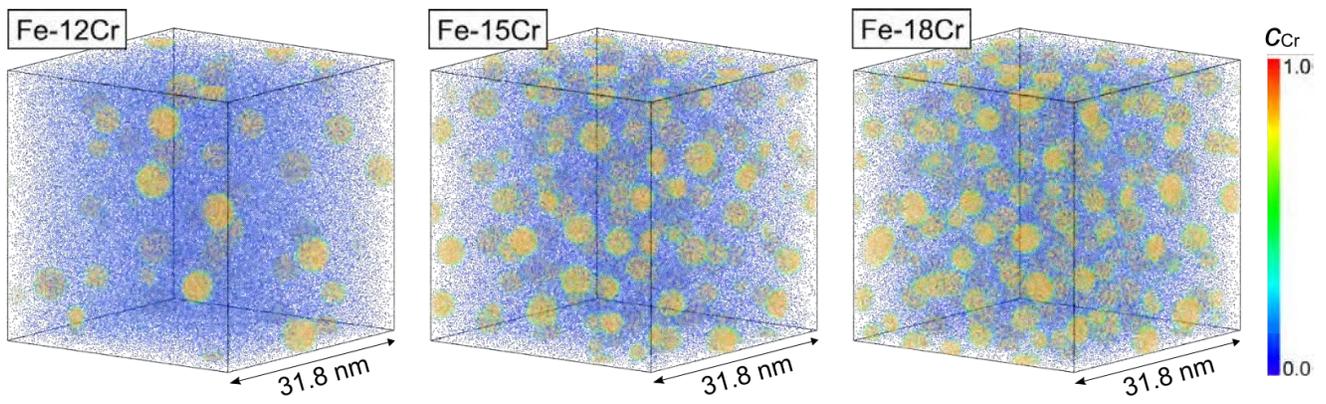

Figure 1. Simulation results showing α' precipitation in neutron-irradiated ($3.4\times10^{-7}$ dpa/s) Fe-12-18 at.% Cr at 320 °C. The color shows the composition of Cr at the local grid points in the 31.8-nm computational supercell.

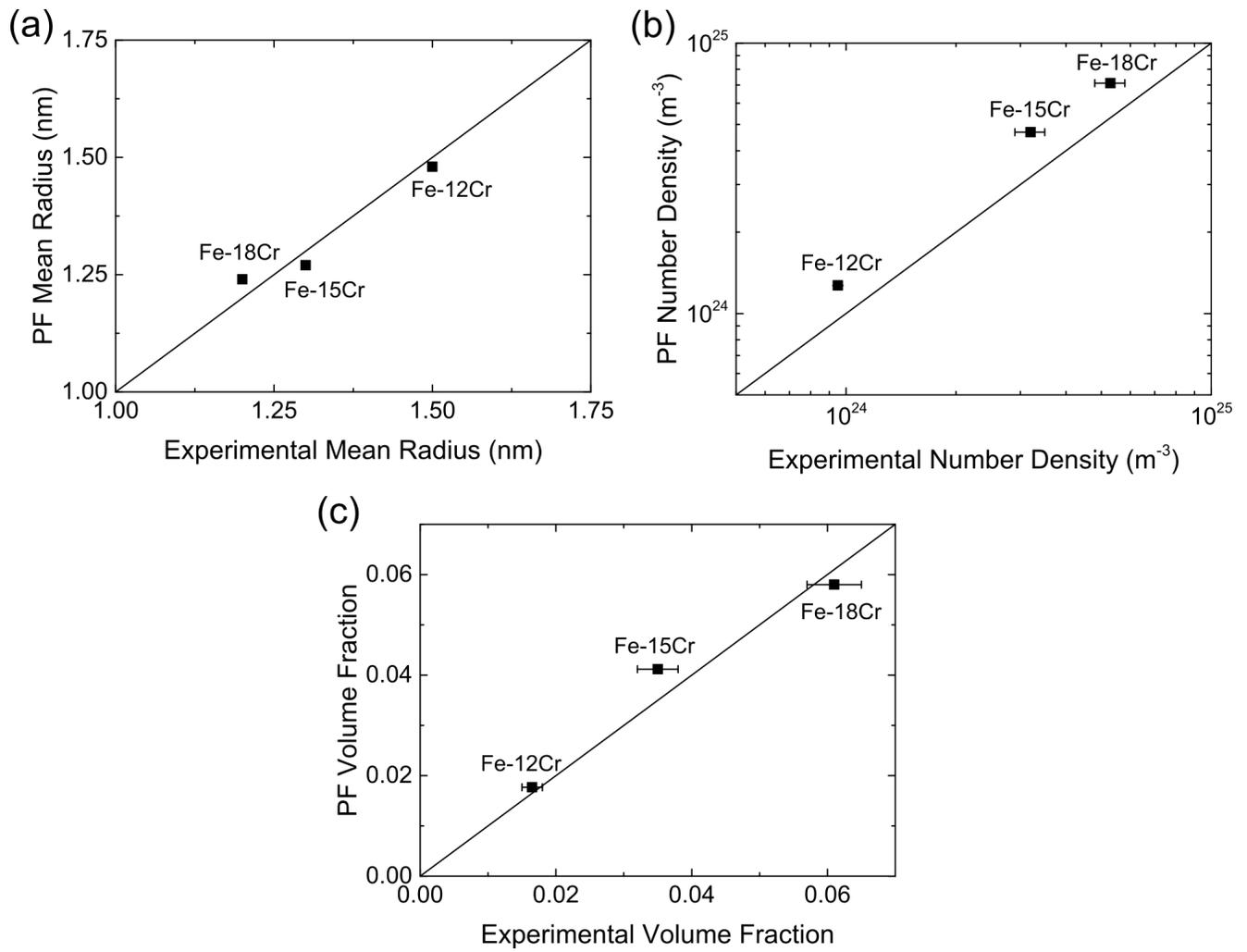

Figure 2. Plots of comparison showing the (a) mean radius, (b) number density, and (c) volume fraction between the phase-field (PF) simulation results and experimental values reported by the APT study [9].

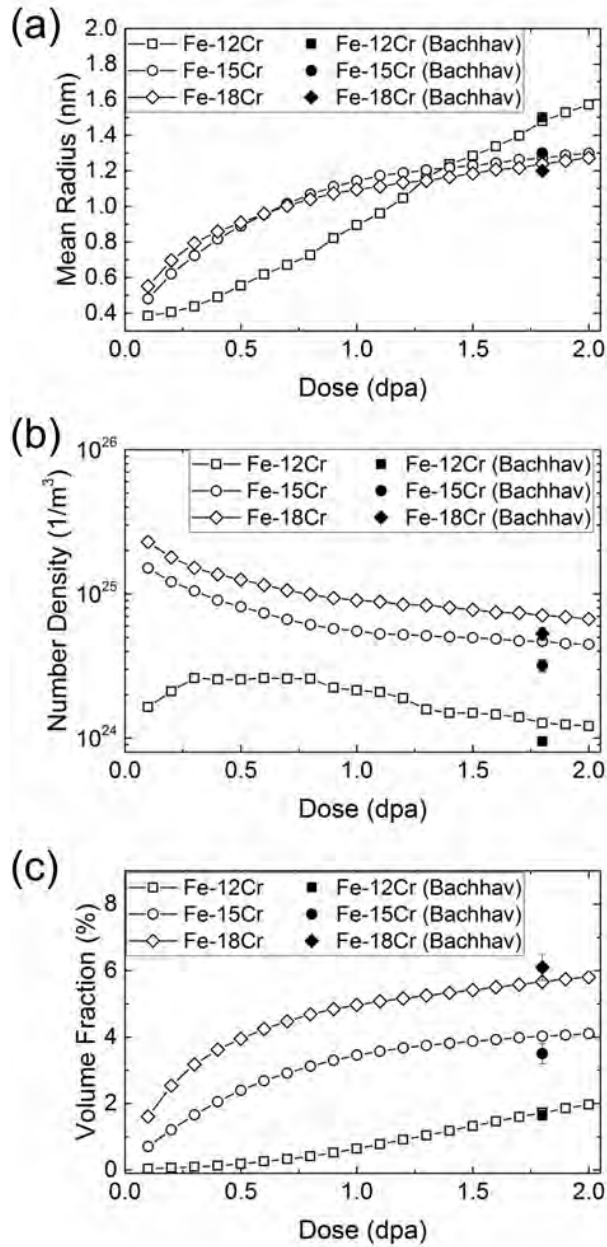

Figure 3. Plots showing the predicted evolution of the (a) mean radius, (b) number density, and (c) volume fraction of α' precipitate in neutron-irradiated ($3.4\times10^{-7}$ dpa/s) Fe-12-18 at.% Cr at 320 °C. The volume fraction is evaluated by using the 80 at.% Cr isosurface. The solid symbols show the experiment measurements [9].

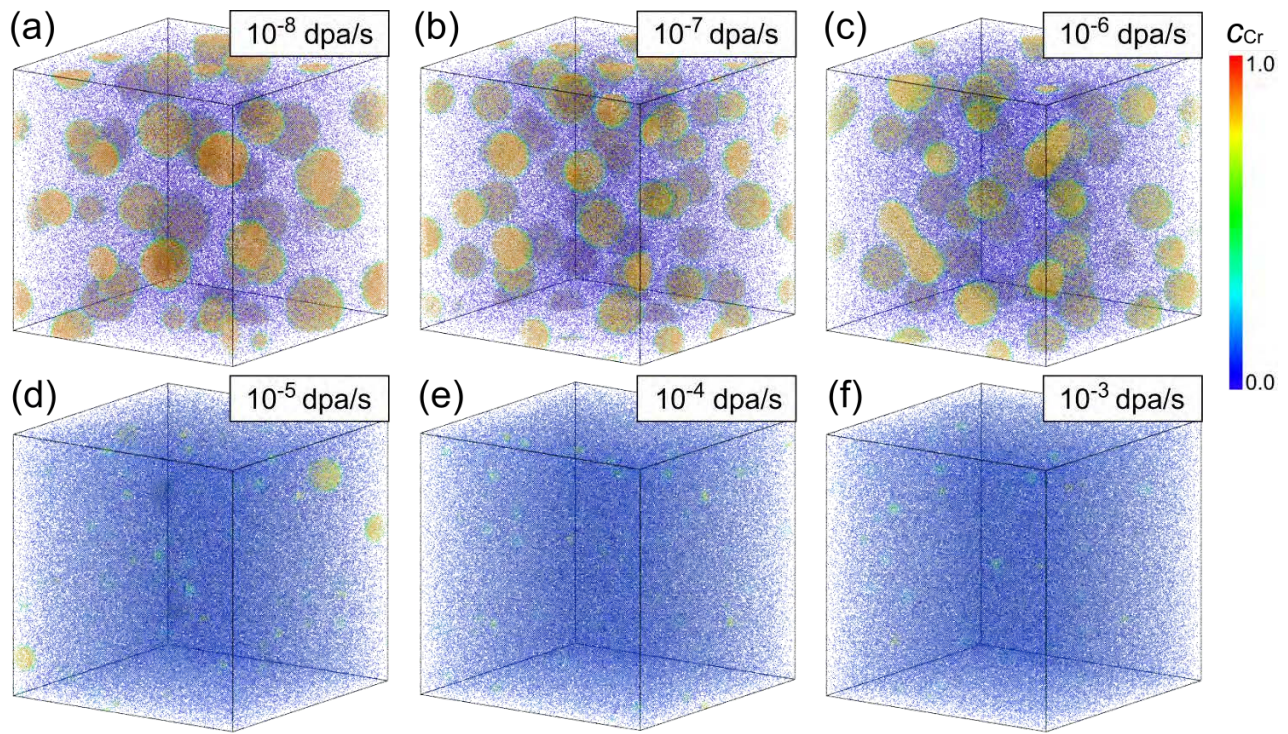

Figure 4. Simulation results showing α' precipitation in Fe-15Cr at 300 °C irradiated to 10 dpa at the dpa rate: (a) $10^{-8}$ dpa/s, (b) $10^{-7}$ dpa/s, (c) $10^{-6}$ dpa/s, (d) $10^{-5}$ dpa/s, (e) $10^{-4}$ dpa/s, and (f) $10^{-3}$ dpa/s. The color shows the composition of Cr at the local grid points in the 31.8-nm computational supercell. Simulations are for neutron or ion irradiation conditions and therefore include strong cascade mixing (see Section 2.4).

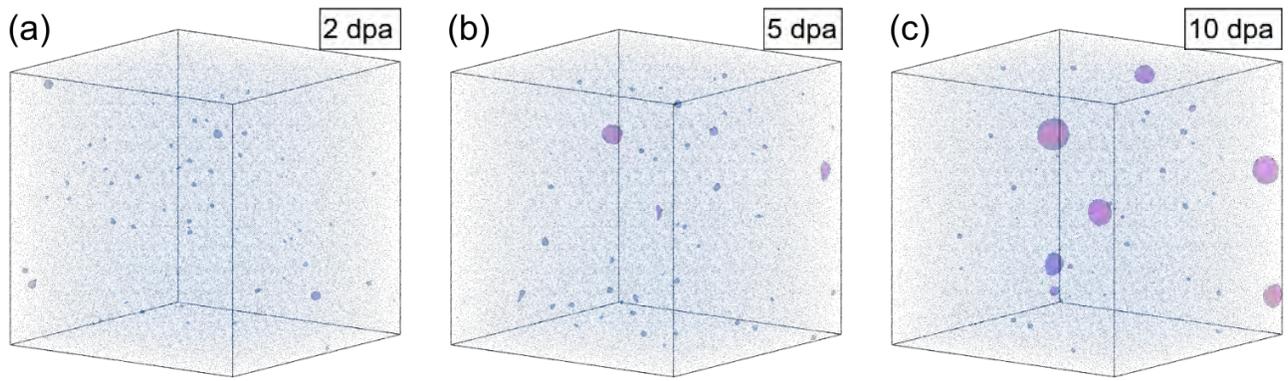

Figure 5. Simulation results showing the evolution of α' precipitation at $10^{-5}$ dpa/s at 300 °C after irradiation to (a)2, (b)5, and (c)10 dpa. The purple color shows the isosurface of 75% Cr. Simulations are for neutron or heavy-ion irradiation conditions and therefore include strong cascade mixing (see Section 2.4).

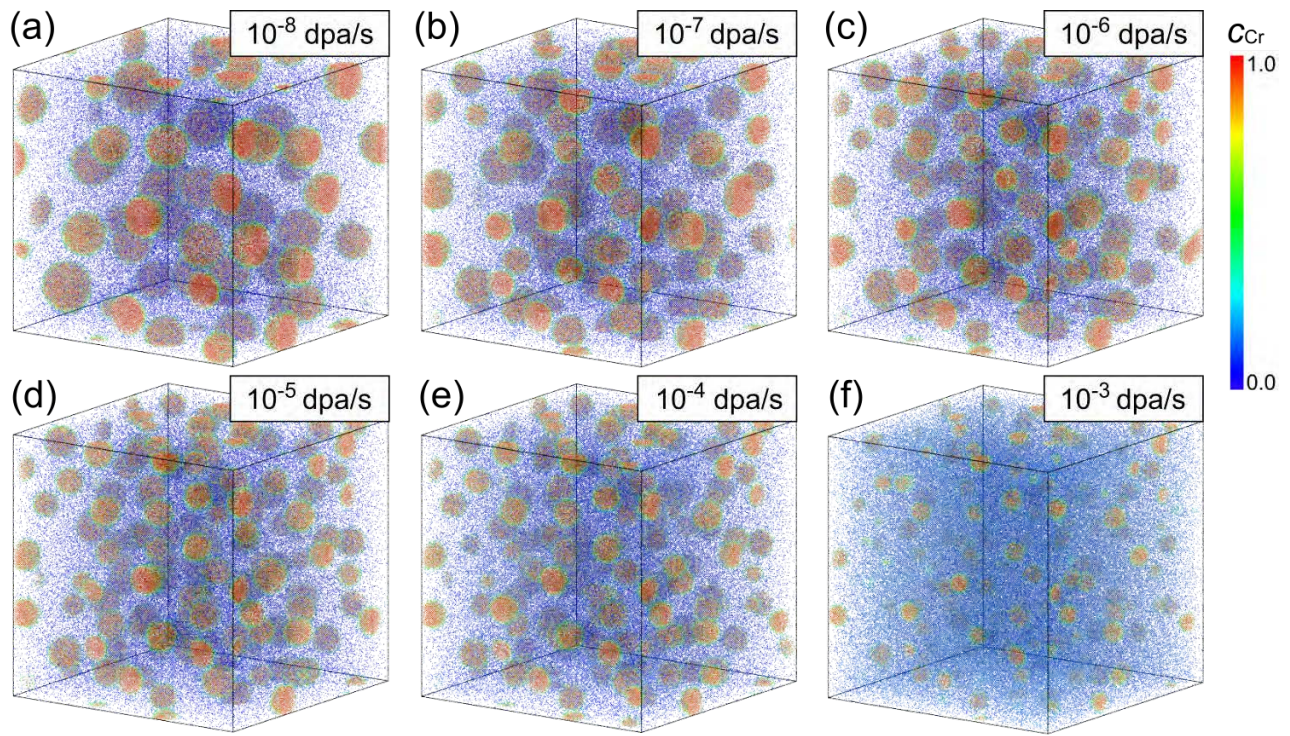

Figure 6. Simulation results showing α' precipitation in Fe-15Cr at 300 °C irradiated to 10 dpa at the dpa rate: (a) $10^{-8}$ dpa/s, (b) $10^{-7}$ dpa/s, (c) $10^{-6}$ dpa/s, (d) $10^{-5}$ dpa/s, (e) $10^{-4}$ dpa/s, and (f) $10^{-3}$ dpa/s. The color shows the composition of Cr at the local grid points in the 31.8-nm computational supercell. Simulations are for electron irradiation conditions with only Frenkel pair production and therefore include weak cascade mixing (see Section 2.4).

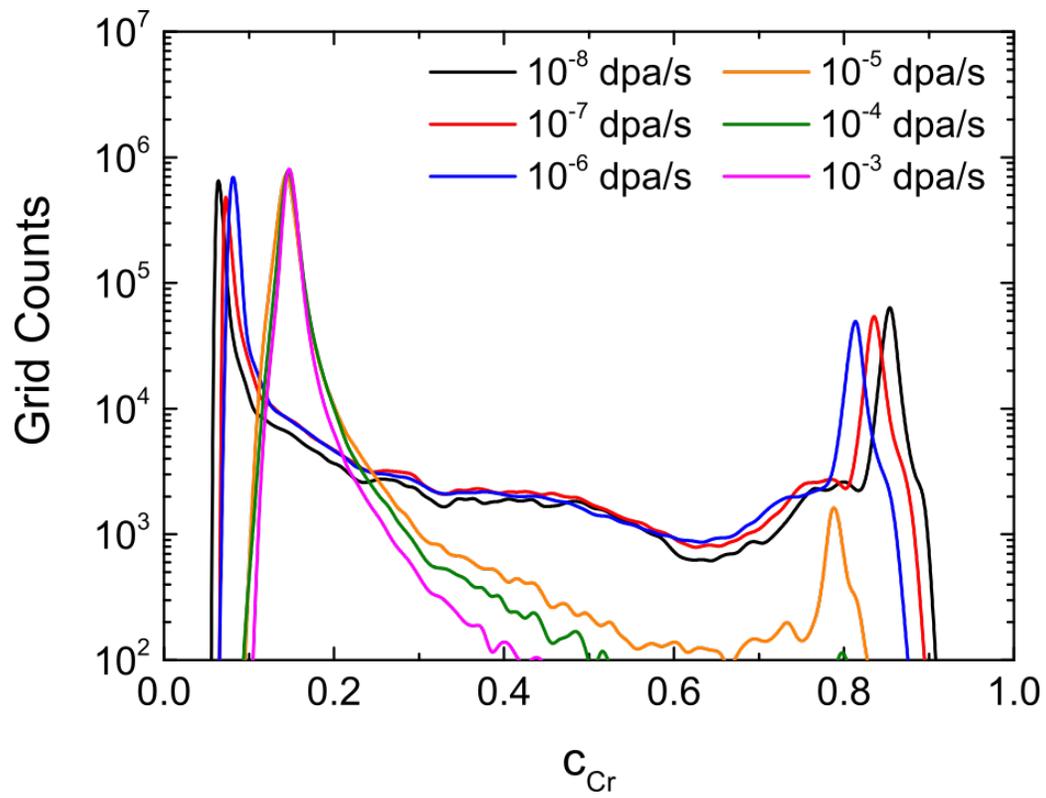

Figure 7. Composition histogram plots showing the Cr concentration in the grid volumes for Fe-15Cr irradiated to 10 dpa at 300 °C (see Figure 4) for dpa rates from $10^{-8}$ to $10^{-3}$ dpa/s. The lines connect points on a histogram with a Cr binning with of 1%.

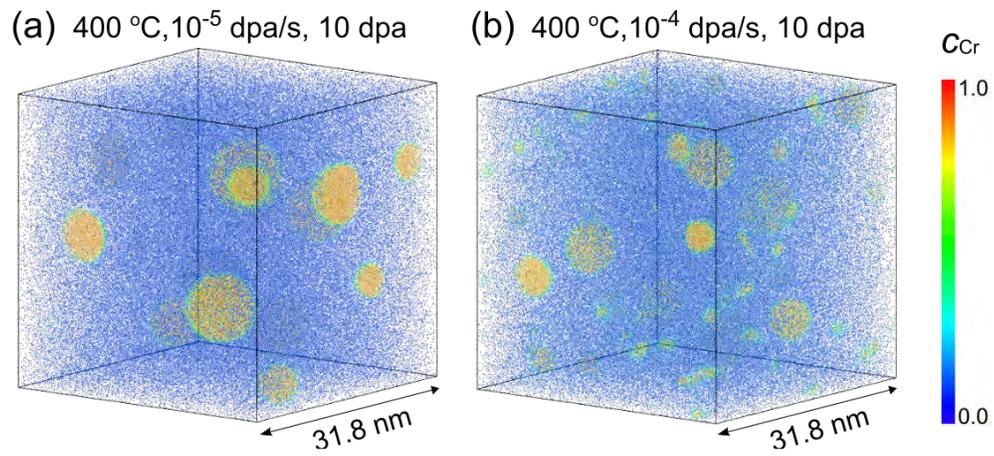

Figure 8. Simulation results showing α' precipitation in Fe-15Cr at 400 °C irradiated to 10 dpa with strong displacement cascades at the dpa rate: (a) $10^{-5}$ dpa/s and (b) $10^{-4}$ dpa/s. The color shows the composition of Cr at the local grid points in the 31.8-nm computational supercell.

# Supplementary Information

**Flux effects in precipitation under irradiation – simulation of Fe-Cr alloys**

Jia-Hong Ke [a], Elaina R. Reese [b], Emmanuelle A. Marquis [b], G. Robert Odette [c], and Dane Morgan [a]

[a] Department of Materials Science and Engineering, University of Wisconsin-Madison, Madison, WI 53706, USA

[b] Department of Materials Science and Engineering, University of Michigan, Ann Arbor, MI 48109, USA

[c] Department of Materials Science and Engineering, University of California, Santa Barbara, CA 93106, USA

* Corresponding author. E-mail address: ddmorgan@wisc.edu

## S1. Gibbs energy of the bcc-α phase in binary Fe-Cr system

The phase diagram of Fe-Cr binary system was assessed by Andersson and Sundman[1]. Table S1 lists the parameters of bcc-α (or α') phase. The Gibbs energy (J m$^{-3}$) of the bcc phase gives

$$f_{ch}^{bcc} = \frac{1}{V_m}\sum_{i=\text{Fe,Cr}} c_i\,^0G_i^{bcc} + RT\sum_{i=\text{Fe,Cr}} c_i \ln c_i + {}^EG^{bcc} + G_m^{bcc} \tag{S1}$$

where $c_i$ is the conserved order parameter or atomic fraction of species $i$, $^0G_i^\rho$ is the Gibbs energy (unit in J/mol) of the pure element $i$ in the chemical phase $\rho$, and $V_m$ is the molar volume. The third term corresponds to the excess Gibbs energy, which can be written by the Redlich-Kister polynomial:

$$^EG^{bcc} = c_{\text{Fe}} c_{\text{Cr}} \sum_{r=0}^{n} \left[ {}^rL_{\text{Cr,Fe}}^{bcc} \left(c_{\text{Cr}} - c_{\text{Fe}}\right)^r \right] \tag{S2}$$

where $^rL_{\text{Fe,Cr}}^{bcc}$ is the temperature-dependent interaction parameters in the binary system. The last term in Eq. **Error! Reference source not found.** corresponds to the magnetic contribution of Gibbs energy, which can be expressed by following the Hillert-Jarl model [2]:

$$G_m^{bcc} = RT \ln(\beta^{bcc}+1) f(\tau) \tag{S3}$$

where $\beta^{bcc}$ is the magnetic moment, $\tau$ is defined as $T/T_C$, $T_C$ is the Curie temperature, and the function $f$ is given by [1]

$$f(\tau) = \begin{cases} -0.9053\tau^{-1} + 1.0 - 0.153\tau^3 - 6.8\times10^{-3}\tau^9 - 1.53\times10^{-3}\tau^{15} & \text{for } \tau < 1 \\ -6.471\times10^{-2}\tau^{-5} - 2.037\times10^{-3}\tau^{-15} - 4.278\times10^{-4}\tau^{-25} & \text{for } \tau > 1 \end{cases} \tag{S4}$$

$\beta^{bcc}$ and $T_C$ can be expressed by the similar polynomial:

$$\beta^{bcc} = c_{Fe}\,{}^0B^{bcc}_{magn\,Fe} + c_{Cr}\,{}^0B^{bcc}_{magn\,Cr} + c_{Cr}c_{Fe}\,{}^0B^{bcc}_{magn\,Cr,Fe} \tag{S5}$$

$$T_C = c_{Fe}\,{}^0T^{bcc}_{C\,Fe} + c_{Cr}\,{}^0T^{bcc}_{C\,Cr} + c_{Cr}c_{Fe}\left({}^0T^{bcc}_{c\,Cr,Fe} + (c_{Cr}-c_{Fe})\,{}^1T^{bcc}_{C\,Cr,Fe}\right) \tag{S6}$$

Table S1. Thermodynamic parameters of the Fe-Cr system [1] utilized in the phase-field model.

| Function | Expression |
|---|---|
| ${}^0G^{bcc}_{Fe}$ | 1224.83 + 124.134×T - 23.5143×T×log(T) - 4.39752×10$^{-3}$×T$^2$ - 5.89269×10$^{-8}$×T$^3$ + 77358.5×T$^{-1}$ |
| ${}^0G^{bcc}_{Cr}$ | -8851.93 + 157.48×T - 26.908×T×log(T) + 1.89435×10$^{-3}$×T$^2$ - 1.47721×10$^{-6}$×T$^3$ + 139250×T$^{-1}$ |
| ${}^0L^{bcc}_{Cr,Fe}$ | 20500 - 9.68×T |
| ${}^0B^{bcc}_{magn\,Fe}$ | 2.22 |
| ${}^0B^{bcc}_{magn\,Cr}$ | -0.008 |
| ${}^0B^{bcc}_{magn\,Cr,Fe}$ | -0.85 |
| ${}^0T^{bcc}_{C\,Fe}$ | 1043 |
| ${}^0T^{bcc}_{C\,Cr}$ | -311.5 |
| ${}^0T^{bcc}_{C\,Cr,Fe}$ | 1650 |
| ${}^1T^{bcc}_{C\,Cr,Fe}$ | 550 |

## S2. Radiation-enhanced diffusion (RED)

We apply the radiation-enhanced diffusion model developed by Odette *et al.* [3] to calculate the vacancy concentration under irradiation and scale thermal diffusion coefficients. By following the general treatment of RED models, the radiation-enhanced atomic mobility coefficient is given as

$$M = X_V^r \frac{M^{th}}{X_V^e} + M^{th} \tag{S7}$$

where $X_V^r$ is the non-equilibrium vacancy concentration under irradiation, $X_V^e$ is the vacancy concentration under thermodynamic equilibrium. $M^{th}$ is the atomic mobility coefficient obtained from the DICTRA database [4]. Under the steady state when defect production is balanced by annihilation at sinks as well as recombination at matrix and solute trapped vacancies, the vacancy concentration can be expressed as [3]

$$X_V^r = \frac{g_s \xi \phi_{dpa}}{D_V S_t} \tag{S8}$$

where $\phi_{\text{dpa}}$ is the dpa rate, $D_V$ is the vacancy diffusion efficient as obtained from Ref. [4], $\xi$ is the cascade efficiency or the fraction of vacancies and self-interstitial atom (SIA) created per dpa, $S_t$ is the total sink strength including the contributions of dislocations ($S_d$) and vacancy clusters ($S_c$). The former can be characterized by the dislocation density and the latter can be evaluated by [3]

$$S_c = 4\pi r_c \phi_{\text{dpa}} \tau_c / V_a \tag{S8}$$

where $r_c$, $\sigma_c$ and $\tau_c$ are the recombination radius, production cross-section and annealing time for vacancy clusters. Note that the spatial saturation of vacancy clusters may occur under high-dpa-rate radiation as the region of radiation damage starts to overlap [5]. Thus we implemented an upper limit of the vacancy cluster density by assuming the exclusion area is in an approximated radius of 10 nm, which results in the saturated sink strength ($S_{sat}$) of vacancy clusters to be $3\times10^{15}$ m$^{-2}$.

The RED model is based on the following two rate theory equations:

$$\frac{\partial X_V}{\partial t} = \xi \phi_{\text{dpa}} + \frac{X_{tv}}{\tau_t} - \frac{4\pi r_r}{V_a}(D_i + D_V)X_i X_V - D_V X_V S_t - \frac{4\pi r_t}{V_a} D_V X_V X_t \tag{S8}$$

$$\frac{\partial X_i}{\partial t} = \xi \phi_{\text{dpa}} - \frac{4\pi r_r}{V_a}(D_i + D_V)X_i X_V - D_i X_i S_t - \frac{4\pi r_t}{V_a} D_i X_i X_{tv} \tag{S8}$$

where $r_r$ and $r_t$ are respectively the matrix and trap recombination radii, $X_t$ is the trap density, $X_{tv}$ is the trap vacancy concentration, $\tau_t$ is the annealing time for trapped vacancies which can be expressed as

$$\tau_t = d^2 \bigg/ D_V e^{-\frac{H_b}{kT}} \tag{S8}$$

where $H_b$ is the binding energy for trapped vacancies and $d$ is the nearest neighbor distance of bcc Fe lattice ($2.48\times10^{-10}$ m). $g_s$ in Eq. (S8) is the vacancy survival fraction which can be obtained by solving the steady-state condition of Eq. (S10) and Eq. (S11) with $X_i D_i = X_V D_V$, which gives [3]:

$$1 - \frac{R_r \xi \phi_{\text{dpa}} (g_s)^2}{D_i D_v S_t^2} - g_s - \frac{\xi \phi_{\text{dpa}} R_t^2 X_t \tau_t (g_s)^2}{S_t^2 (1 + g_s \xi \phi_{\text{dpa}} R_t \tau_t / S_t)} = 0 \tag{S8}$$

Here we consider the highly concentrated Cr atoms in 12-18Cr alloys as the main trapping site, so the trap density $X_t$ is the Cr composition in the model alloy. $D_V$ is approximated by $10^{-4} \exp(-E_m/kT)$ where $E_m$ is the migration energy of vacancies [6]. The vacancy migration energies in dilute bcc-α solid solution and that of bcc-α' were reported respectively as 1.1 eV and 1.3 eV [7,8]. In the present study we roughly assume that the vacancy migration energy follows the linear relationship with respect to the Cr content. The calculation result of the RED model is provided in Figure S1 that shows the radiation-enhanced diffusion coefficient of Cr in Fe-12Cr, Fe-15Cr, and Fe-18Cr alloys as a function of dose rates at 300 °C. Note that the cusp in Figure S1 is caused by the spatial saturation of vacancy clusters as implemented in the RED model.

Table S2 lists the parameters used in the present study with references. Note that vacancy cluster recombination radius and vacancy cluster annealing time are treated as fitting parameters to match the neutron data of precipitation. The vacancy cluster recombination radius typically has a value on the scale of a few nearest-neighbor interatomic distances, which distance for Fe is 0.248 nm. Our fitting vacancy cluster recombination radius of 0.30 nm is therefore very physically reasonable. The vacancy cluster annealing time depends strongly on temperature and vacancy cluster binding energy and is given by $8.1 \times 10^{-12}/\exp(-E_{VB}/kT)$ [10], where $E_{VB}$ is the vacancy cluster binding energy. $E_{VB}$ = 1.855 eV was used in a previous study [10]. In the present work we treated $E_{VB}$ as a fitting parameter by adjusting the value in the range 1.85 ± 0.3 eV to find the best fit to neutron data. $E_{VB}$ = 1.70 eV was finally fitted and used in this simulation study. This value is well within the uncertainty of this quantity under different alloys and conditions and therefore quite reasonable.

Table S1. Parameters used in the RED model for calculating radiation-enhanced diffusion. * The variables marked with a *, which include "Vacancy cluster recombination radius" and "Vacancy cluster annealing time" that were fitted to match the size, number density and volume fraction of α'

precipitates after 1.8 dpa of neutron irradiation at 320 °C [9].

| | | |
|---|---|---|
| *Vacancy cluster recombination radius | $r_c$ | $3.0\times10^{-10}$ m [10] |
| *Vacancy cluster annealing time | $\tau_c$ | $8.1\times10^{-12}/e^{(-1.70/kT)}$ [10] |
| Vacancy cluster production cross-section | $\sigma_c$ | $1.5\times10^{-29}$ m² [3] |
| Cascade efficiency | $\xi$ | 0.33 (neutron or ion irradiation) [11]<br>1.0 (electron) [11] |
| Atomic volume | $V_a$ | $1.18\times10^{-29}$ m³ |
| Vacancy migration energy | $E_m$ | 1.1 eV (for dilute Fe-Cr) [7]<br>1.3 eV (for pure Cr) [8] |
| Dislocation sink strength (dislocation density) | $S_d$ | $1\times10^{13}$ m⁻² |
| Matrix recombination radius | $r_r$ | $5.7\times10^{-10}$ m [3] |
| Trap recombination radius | $r_t$ | $5.7\times10^{-10}$ m [3] |
| Binding energy for trapped vacancies | $H_b$ | 0.094 eV [12] |
| Saturated sink strength of vacancy clusters | $S_{sat}$ | $3\times10^{15}$ m⁻² |

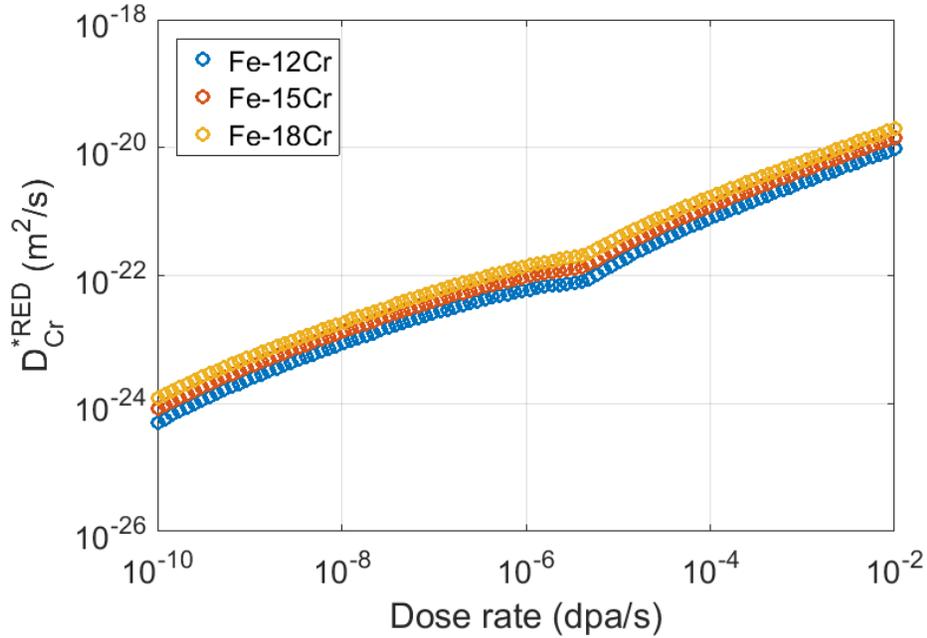

Figure S1. Plots showing the radiation-enhanced diffusion coefficient of Cr in Fe-12Cr, Fe-15Cr, and Fe-18Cr alloys as a function of the dpa rate at 300 °C.

## S3. Dose-evolution of α' size distribution in neutron-irradiated Fe-Cr alloys at 320 °C

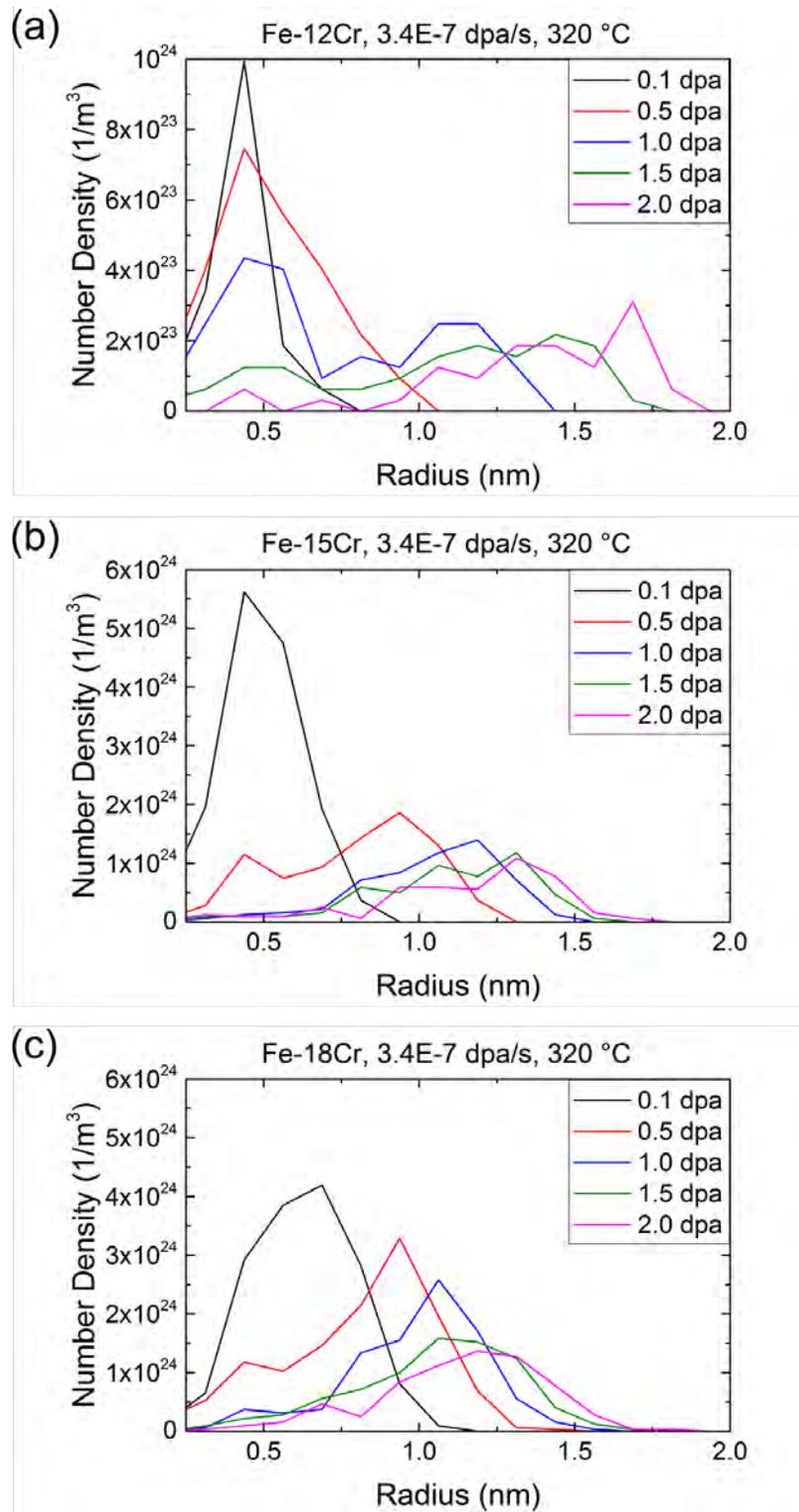

Figure S2. Simulation results showing the evolution of α' precipitate size distribution in (a) Fe-12Cr, (b) Fe-15Cr, and (c) Fe-18Cr after neutron irradiation at 320 °C to 0.1, 0.5, 1.0, 1.5, and 2.0 dpa.

## S4. Dpa dependence of α' number density, mean radius, volume fraction, α and α' Cr fraction at different dose rates for cascade-mixing neutron/heavy-ion irradiation.

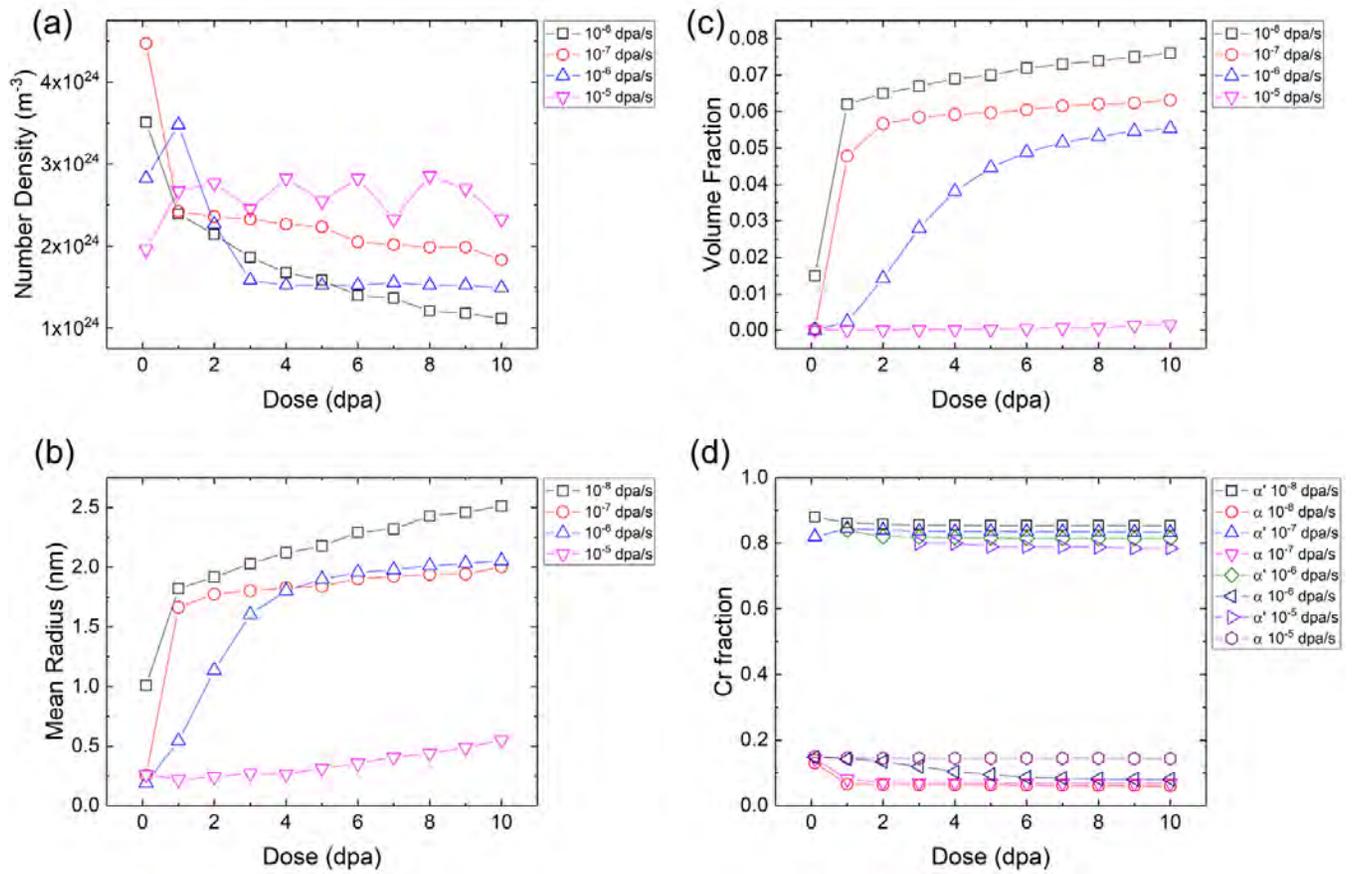

Figure S3. Plots showing the (a) α' number density, (b) mean radius, (c) volume fraction, (d) α and α' Cr fraction as a function of dpa at 300 °C under neutron/heavy-ion irradiation at various dpa rates. The size and volume fraction are evaluated by using the 80 at.% Cr isosurface. Note that for $10^{-5}$ dpa/s 75 at.% is used due to the lower α' Cr composition than 80 at.%.

## S5. Dpa dependence of α' number density, mean radius, volume fraction, α and α' Cr fraction at different dose rates for electron irradiation with cascade mixing allowed.

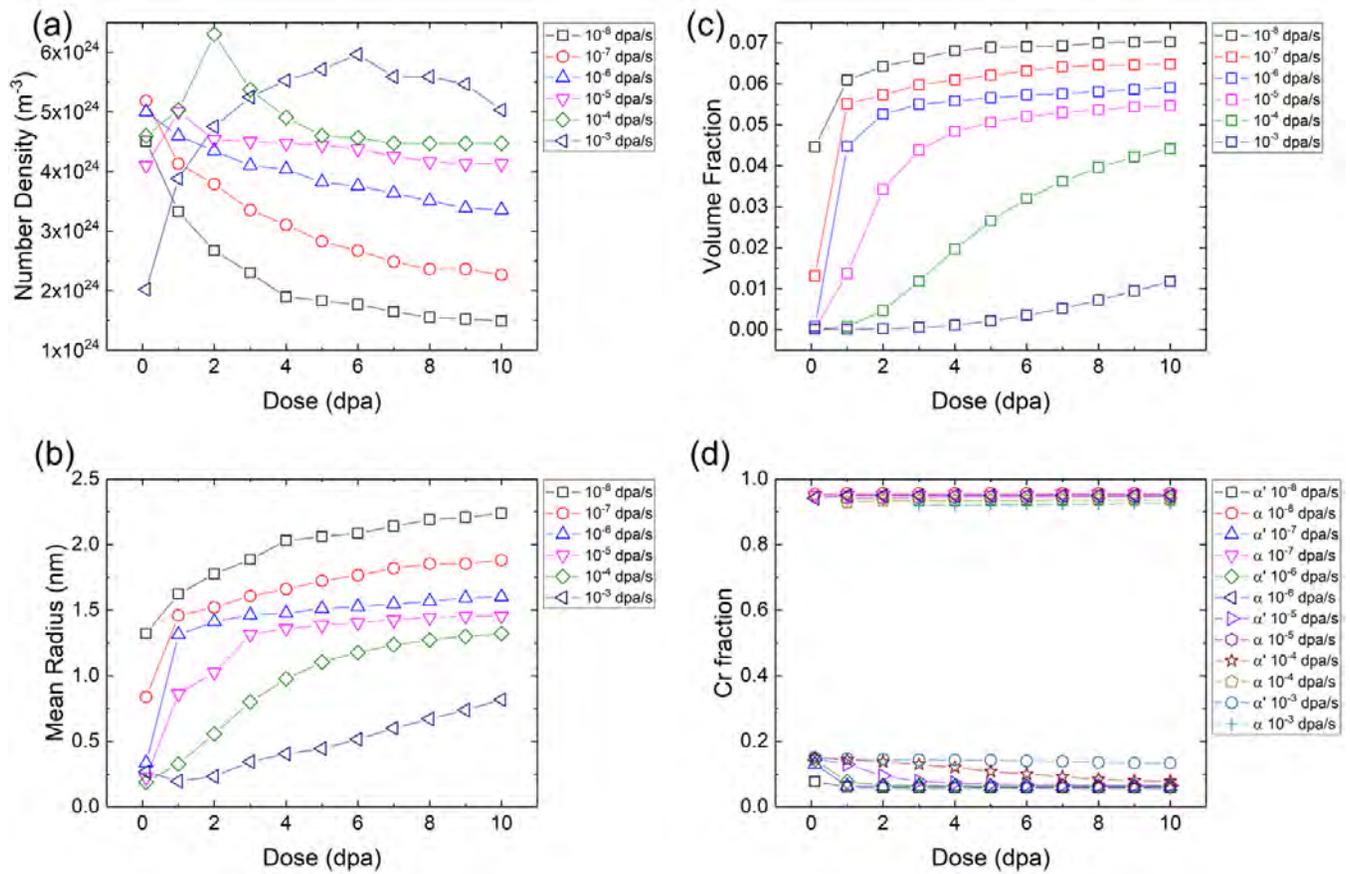

Figure S4. Plots showing the (a) α' number density, (b) mean radius, (c) volume fraction, (d) α and α' Cr fraction as a function of dpa at 300 °C under electron irradiation at various dpa rates. The size and volume fraction are evaluated by using the 80 at.% Cr isosurface.

## S6. Dpa dependence of α' number density, mean radius, volume fraction, α and α' Cr fraction at different dose rates for electron irradiation without cascade mixing.

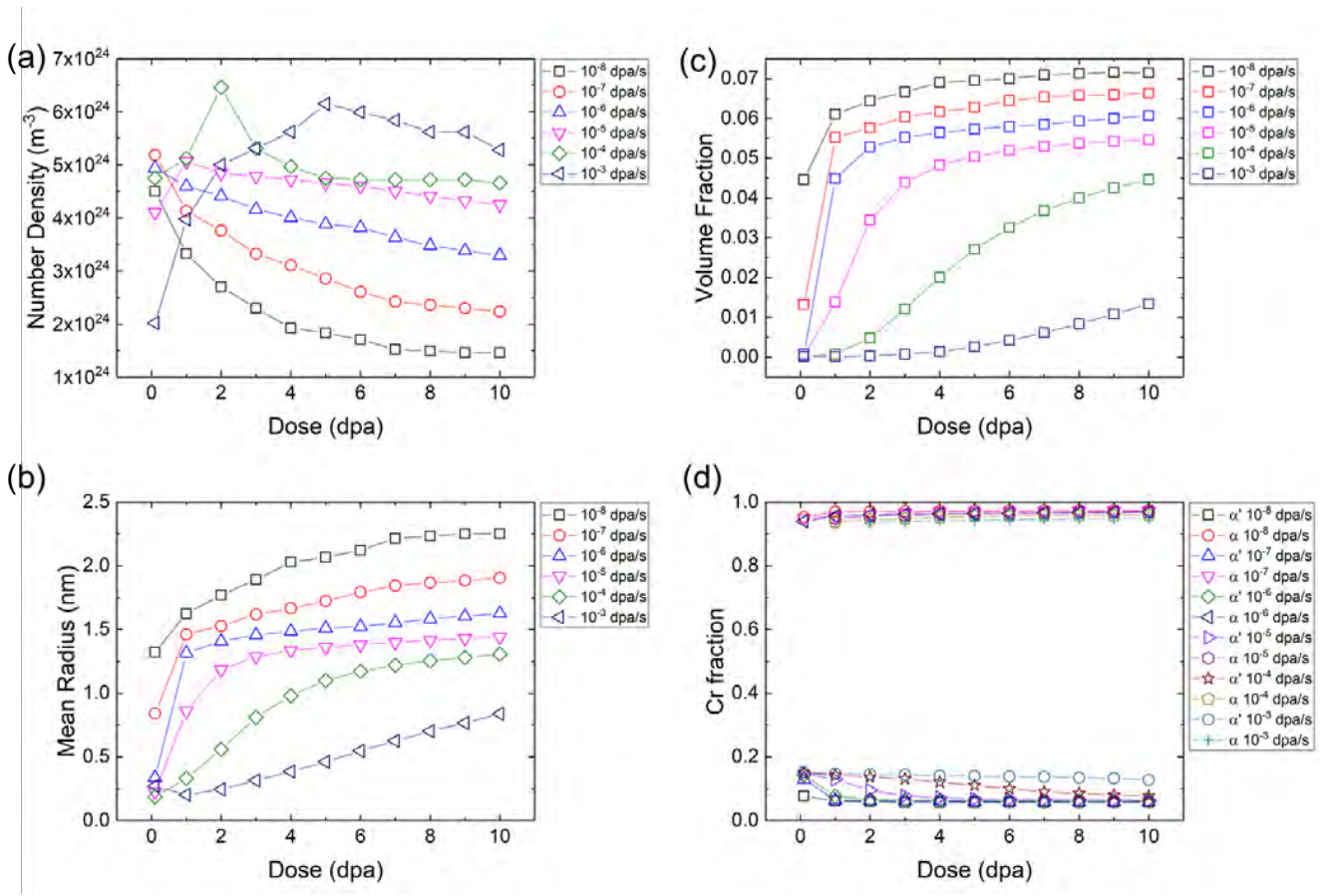

Figure S5. Plots showing the (a) α' number density, (b) mean radius, (c) volume fraction, (d) α and α' Cr fraction as a function of dpa at 300 °C under electron irradiation without cascade mixing at various dpa rates. The size and volume fraction are evaluated by using the 80 at.% Cr isosurface.

## S7. Cr composition profile (10 dpa, 300°C under electron irradiation at $10^{-6}$ dpa/s)

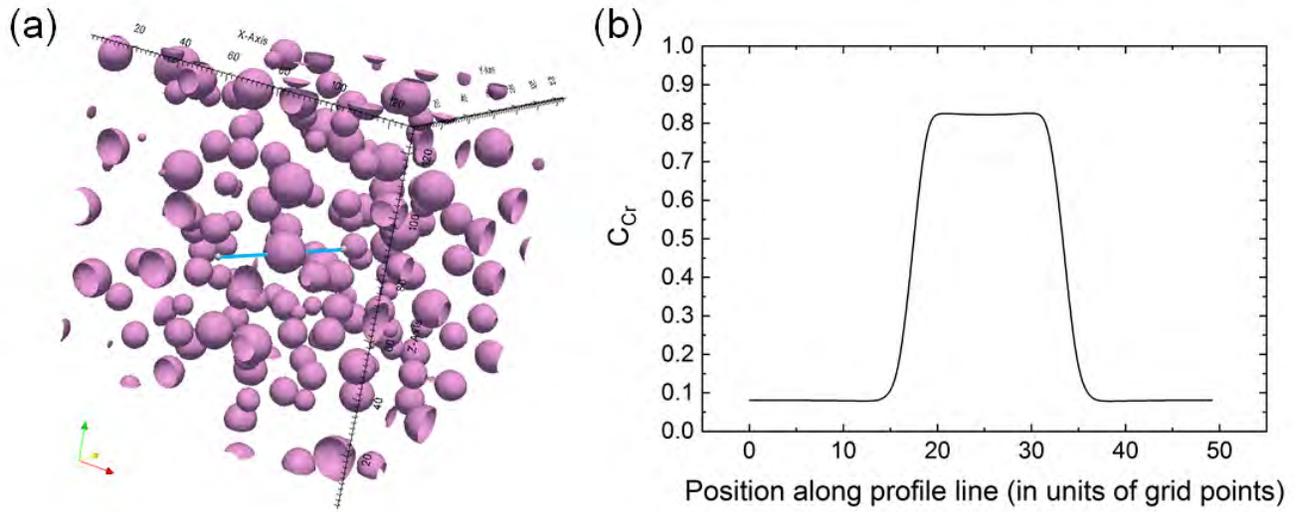

Figure S6. Plots showing (a) the microstructure (80% Cr isosurface) of Fe-15Cr irradiated to 10 dpa at 300°C under electron irradiation at $10^{-6}$ dpa/s and (b) the Cr composition profile along the blue line shown in (a). The peak (position ≈18 - ≈33 in (b)) has a large region of approximately 82% Cr and clearly shows a stable particle, which can also be see in (a). One unit of grid length is 0.249 nm.